\documentclass[twocolumn,showpacs,
  aps,superscriptaddress,
  prd,notitlepage,showkeys,
  nofootinbib]{revtex4-1}
 
\usepackage[normalem]{ulem}
\usepackage{amssymb}
\usepackage{amsmath}
\usepackage{graphicx,subfigure}
\usepackage{dcolumn}
\usepackage[colorlinks,urlcolor=blue,citecolor=blue,linkcolor=Maroon]{hyperref}
\usepackage{color,units}
\usepackage[dvipsnames]{xcolor} 
\usepackage{lineno}
\usepackage{xspace}
\usepackage{longtable} 
\usepackage{float}  
\usepackage{aas_macros}
\usepackage{amsfonts,wasysym,epsfig,verbatim,subfigure,bm,mathrsfs,lipsum}
\usepackage{comment}
\usepackage{braket}
\usepackage{physics}
\usepackage{hyperref}

\begin{document}

\newcommand{\IUCAA}{Inter-University Centre for Astronomy and
  Astrophysics, Post Bag 4, Ganeshkhind, Pune - 411007, India}
  
\newcommand{\NIKHEF}{Nikhef - National Institute for Subatomic Physics, Science Park, 1098 XG Amsterdam, Netherlands}

\newcommand{\UVA}{Institute for High-Energy Physics, University of Amsterdam, Science Park, 1098 XG Amsterdam, Netherlands}

\newcommand{\WSU}{Department of Physics and Astronomy, Washington State University, 1245 Webster, Pullman, Washington 99164-2814, U.S.A.}

\newcommand{\GRAPS}{Institute for Gravitational and Subatomic Physics, Utrecht University, Princetonplein 1, 3584 CC Utrecht, Netherlands}

\newcommand{\MPI}{Max-Planck-Institut f{\"u}r Gravitationsphysik (Albert-Einstein-Institut), D-30167 Hannover, Germany}

\newcommand{\LBNZ}{Leibniz Universit{\"a}t Hannover, D-30167 Hannover, Germany}

\newcommand{\sukanta}[1]{{\color{magenta}\bf  SB:} {\color{magenta} #1}}
\newcommand{\sayak}[1]{{\color{red}\bf  SD:} {\color{red} #1}}
\newcommand{\samanwaya}[1]{{\color{ForestGreen}\bf  SM:} {\color{ForestGreen} #1}}

\newcommand{\new}[1]{\textcolor{blue}{#1}}

\title{Towards establishing the presence or absence of horizons in coalescing binaries of compact objects by using their gravitational wave signals
}
\author{Samanwaya Mukherjee}
\email{samanwaya@iucaa.in}
\affiliation{\IUCAA}

\author{Sayak Datta}\email{sayak.datta@aei.mpg.de}
\affiliation{\IUCAA}
\affiliation{\MPI}
\affiliation{\LBNZ}

\author{Srishti Tiwari}
\affiliation{\IUCAA}

\author{Khun Sang Phukon}
\affiliation{\NIKHEF}
\affiliation{\UVA}
\affiliation{\GRAPS}

\author{Sukanta Bose}
\affiliation{\IUCAA}
\affiliation{\WSU}



    \begin{abstract}
      The quest for distinguishing black holes from horizonless compact objects using gravitational wave signals from coalescing compact binaries can be helped by utilizing the phenomenon of \textit{tidal heating}, which leaves its imprint on the binary waveforms through the \textit{horizon parameters}. These parameters, defined as $H_1$ and $H_2$ with $H_{1,2} \in [0,1]$ for the two compact objects, are combined with the binary components' masses and spins to form two new parameters, $H_{\rm eff5}$ and $H_{\rm eff8}$, to minimize their covariances in parameter estimation studies. In this work, we investigate the effects of tidal heating on gravitational waves to probe the observability of these effective parameters. We use a post-Newtonian waveform that includes the phase contribution due to tidal heating as a function of $H_{\rm eff5}$ and $H_{\rm eff8}$, and examine their 1-$\sigma$ measurement errors as well as the covariances between them mainly using the \textit{Fisher matrix} approach. Since this approach works well for high signal-to-noise ratios, we focus primarily on the third generation (3G) gravitational wave detectors Einstein Telescope and Cosmic Explorer and use the second generation (2G) detector-network of LIGO (Hanford, Livingston) and Virgo for comparison. We study how the errors vary with the binaries' total mass, mass-ratio, luminosity distance, and component spins.  We find that the region in the total binary mass where measurements of $H_{\rm eff5}$ and $H_{\rm eff8}$ are most precise are $\sim 20 - 30M_\odot$ for LIGO-Virgo and $\sim 50 - 80M_\odot$ for 3G detectors. Higher component spins allow more precise measurements of $H_{\rm eff5}$ and $H_{\rm eff8}$. For a binary situated at 200 Mpc with component masses $12M_\odot$ and $18M_\odot$, equal spins $\chi_1=\chi_2=0.8$, and $H_{\rm eff5}=0.6$, $H_{\rm eff8}=12$, the 1-$\sigma$ errors in these two parameters are $\sim 0.01$ and $\sim 0.04$, respectively, in 3G detectors. These estimates suggest that precise measurements of the horizon parameters are possible in third-generation detectors, making tidal heating a potential tool to identify the presence or absence of horizons in coalescing compact binaries. We substantiate our results from Fisher studies with a set of Bayesian simulations. 
    \end{abstract}
  
\maketitle


\section{Introduction}
\label{intro}

Detection of gravitational waves (GWs) from coalescence of numerous compact binaries by LIGO~\cite{TheLIGOScientific:2014jea} and Virgo~\cite{TheVirgo:2014hva} has opened up a new era of astronomy~\cite{LIGOScientific:2018mvr, LIGOScientific:2020ibl}. Their observations have motivated a series of tests of General Relativity (GR)~\cite{LIGOScientific:2019fpa,Abbott:2018lct}. The components of the binaries observed by LIGO and Virgo are mainly inferred to be either black holes (BHs) or neutron stars (NSs), which is primarily based on the measurements of component masses, population models, and tidal deformability of NSs~\cite{Cardoso:2017cfl}. 
Merger of two NSs was observed in the event GW170817~\cite{gw170817}, and possibly also GW190425~\cite{Abbott:2020uma}. More recently, confirmed detections of events GW200105 and GW200115~\cite{bhns_LIGOScientific:2021qlt} were made where one of the components is believed to be a BH, and the other an NS. However, for the heavier LIGO-Virgo binaries~\cite{LIGOScientific:2018mvr}, it remains to be conclusively proven whether their components are, in fact, BHs of GR or not. 

Indeed, there are various types of exotic compact objects (ECOs)~\cite{Yunes:2016jcc,Cardoso:2016oxy,Aneesh:2018hlp} proposed that are claimed to mimic BHs in these binaries. However, incorrect inferences about their true nature can have far-reaching implications, such as on population models of compact objects.
In the future, the proposed 3rd generation ground-based GW detectors Einstein Telescope~\cite{maggiore2020science} and Cosmic Explorer~\cite{Reitze:2019iox} are expected to have order-of-magnitude better sensitivity compared to current ones in estimating the source parameters, which should enable us to probe the nature of these objects more accurately. Multiple models of ECOs have been proposed, including 
Planck-scale modifications of BH horizons~\cite{Lunin:2001jy, Almheiri:2012rt}, gravastars,~\cite{Mazur:2004fk}, and boson stars~\cite{Liebling:2012fv}, among others. Building separate models for each of these exotic objects is a hard problem, and accurate measurements of their properties are not yet possible with the current detectors. So, a more practical approach would be to devise tests that are generic and
model-independent and are based on our understanding of binary black hole (BBH) dynamics. 

One way to probe the presence of ECOs against BHs is to understand the possible ways in which the characteristics of these objects can differ from those of BHs, and that can be
confirmed or ruled out by introducing appropriate free parameters in the gravitational waveform. In order to develop such model-independent tests of BH mimickers, it is
important to identify the properties that are unique to BHs, and investigate their imprints on the gravitational waveform so that we can measure them from observations.   

Several tests have been proposed to probe whether the compact objects in a binary are BHs or ECOs. One of them is using {\it echoes} to distinguish the remnants of binary merger from BHs, which has initiated rigorous modelling and search for those features in GW data~\cite{Cardoso:2016rao, Cardoso:2016oxy, Abedi:2016hgu, Westerweck:2017hus, Conklin:2017lwb, Maggio:2019zyv, Conklin:2019fcs, Tsang:2019zra, Cardoso:2019rvt, Chen:2020htz, Holdom:2019bdv, Xin:2021zir, Ren:2021xbe, Srivastava:2021uku}. Construction of waveforms for binary ECOs has also begun \cite{Toubiana:2020lzd, Bezares:2022obu}. Measurement of tidal deformability (TD)~\cite{Cardoso:2017cfl, Sennett:2017etc, Maselli:2017cmm, Datta:2021hvm} and spin-induced multipole moments \cite{Krishnendu:2017shb, Datta:2019euh, Bianchi:2020bxa, Mukherjee:2020how, Datta:2020axm, Saleem:2021vph, Krishnendu:2019tjp, Johnson-Mcdaniel:2018cdu} from the late inspiral phase can also be used to test the presence of BHs. Nature of the compact objects can be probed with GWs emitted due to superradiant instability as well~\cite{Barausse:2018vdb,Choudhary:2020pxy}.

Due to their causal structure, BHs in GR are perfect absorbers that behave as dissipative systems~\cite{MembraneParadigm,Damour_viscous,Poisson:2009di,Cardoso:2012zn}.  A significant feature of a BH is its horizon, which is a null surface and a ``one-way membrane" that does not allow energy to escape outward. These tidal effects cause changes in their mass, angular momentum, and horizon area. This phenomenon is called {\it tidal heating} (TH)~\cite{Hartle:1973zz,Hughes:2001jr,PoissonWill}. If the BH is nonspinning, then energy and angular momentum can only flow into the BH. However, spinning BHs can transfer their rotational energy from the ergoregion out into the orbit due to tidal interactions with their binary companion.
Energy exchange via  TH backreacts on the binary's evolution, resulting in a shift in the phase of the GWs emitted by the system. 
Effect of TH on objects such as NSs or horizonless ECOs is comparatively much less due to their lack of a horizon. So, a careful measurement of this phase shift can be used
 in principle to distinguish BHs from horizonless compact objects~\cite{Maselli:2017cmm, Datta:2019euh, Datta:2019epe, Datta:2020rvo, Agullo:2020hxe, Chakraborty:2021gdf, Sherf:2021ppp, Datta:2021row, Sago:2021iku, Maggio:2021uge, Sago:2022bbj}.

To quantify this effect, two ``horizon parameters", $H_1$ and $H_2$, were introduced in a recent study~\cite{datta2020recognizing} whose utility we will study further in characterizing compact objects. These parameters take the value of 1 when the objects are BHs, and $0\leq H_1,H_2<1$ for other compact objects. The phase shift in GWs due to TH will depend on these parameters. Their accurate measurement, in turn, will indicate the presence or absence of BHs in a binary. It turns out that the covariance of these two parameters is generally finite. We therefore find two other related parameters that are mostly statistically independent. Even then, some covariances between the new parameters can arise due to waveform systematics, non-stationary detector noise, etc. Parameter estimation (PE) exercises for real GW signals widely use Bayesian approaches~\cite{datta2020recognizing, Thrane_2019}, which is a robust method, but computationally expensive. Fisher studies~\cite{Vallisneri:2007ev} can provide reliable estimates for the errors and uncertainties in measuring the source parameters of GW signals with high signal-to-noise ratios, and is much faster and less expensive. In the current work, we will primarily use the latter approach for estimating the errors, and explore Bayesian simulations to corroborate the results.

A compact binary coalescence (CBC) consists of three major phases - inspiral, merger, and ringdown. One can model the inspiral phase using post-Newtonian (PN) formalism, whereas numerical relativity (NR) simulations are needed to model the merger regime~\cite{Pretorius:2007nq}. In order to study the ringdown part of the dynamics, one may use BH perturbation theory techniques~\cite{Sasaki:2003xr} or NR. Tidal heating is relevant in the inspiral and is more significant for a binary when the components are closer together so that their tidal interactions are stronger. In the PN regime, TH can be incorporated into the gravitational waveform by adding the phase shift due to this effect into a PN approximant in the time or frequency domain. 

In Sec.~\ref{theory_TH}, we will review the basic framework of tidal heating, describing the waveform parameterization chosen for this work. Sec.~\ref{fisher_basics} will summarize the concepts of the Fisher matrix analysis, discussing various relevant aspects of it, and the corresponding results will be presented in Sec.~\ref{results}. In Sec.~\ref{sec:bayesian} we will present results from  Bayesian simulations to check the consistency of our analyses. In Sec.~\ref{pca}, we will discuss the covariances between the relevant parameters and possibilities of improving the results by diagonalizing the Fisher matrix. We will summarize the results in Sec.~\ref{discussion}, and discuss the relevance of this work to future studies.

Throughout the article, we will use geometric units, assuming $G=c=1$, except when calculating physical quantities.


\section{Theory of Tidal Heating and waveform parameterization}
\label{theory_TH}

The PN formalism~\cite{Blanchet:2013haa} describes the gravitational waveform emitted by a stellar-mass compact binary in its early inspiral phase. In this formalism, the evolution of the orbital phase $\Psi(t)$ of a compact binary is computed as a perturbative expansion in a small parameter, typically taken to be the characteristic velocity $v = (\pi M f)^{1/3}$. Here $M$ is the total mass of the binary and $f$ is the instantaneous GW frequency. This analytical procedure demands $v\ll 1$, which makes it useful in the early inspiral phase of a CBC. For building a proper PN waveform, one begins with the gravitational waveform from an inspiraling pair of point particles (PP). Extra terms are added based on the nature of the binary components. If a component has a finite size and inner structure, e.g., an NS, then {\it tidal deformability}  plays an important role \cite{Flanagan:2007ix}. If there is a BH involved, then the effect of its horizon has to be considered. This is where {\it tidal heating} comes into play.

An electrically neutral spinning black hole, called a Kerr black hole (KBH) in GR parlance, is stationary when it is isolated. On the other hand, when a KBH is a member of a binary, it feels its companion's tidal field, which acts as a non-axisymmetric perturbation~\cite{Hartle:1973zz}. This perturbation causes changes in the mass, spin, and horizon area of the KBH over time~\cite{Alvi:2001mx}. Since the KBH experiences the tidal field of its orbiting companion, it absorbs (emits) energy from (into) the orbit. The absorption part is present in non-spinning BHs as well. Additionally, for a KBH, the difference between the spin frequency and the angular frequency of the tidal field causes the spin to slow down, which in turn makes the KBH lose its rotational energy. The slowing down of a rotating BH due to the gravitational dissipation produced by exterior mass is analogous to the slowing down of a rotating planet by viscous dissipation due to tides raised by an exterior moon that increases its internal thermal content - a phenomenon known as tidal heating. Due to this similarity, the energy and angular momentum flux in BBHs is also termed tidal heating~\cite{Poisson:2004cw}.

The gravitational waveform for a specific binary will include contributions from these factors depending on its components. For a generic binary, we can write the frequency domain strain $\Tilde{h}(f)$ as 

\begin{equation}
\label{hf}
    \Tilde{h}(f) =  \Tilde{A}(f) e^{i\left(\Psi_{\rm PP}+\Psi_{\rm TD}+\Psi_{\rm TH}\right)},
\end{equation}
where $\Tilde{A}(f)$ is the frequency-dependent amplitude. The phase terms -- $\Psi_{\rm PP}, \Psi_{\rm TD},$ and $\Psi_{\rm TH}$ -- arise from the point-particle approximation, tidal deformability, and tidal heating, respectively.

Since GW absorption is negligible for matter~\cite{Glampedakis:2013jya},
TH can be a way to discern the existence of horizons~\cite{Datta:2019euh, Maselli:2017cmm}. Reference~\cite{Datta:2019euh} introduced the {\it horizon parameter} $H$ for extreme mass-ratio inspirals (EMRIs) for this purpose. In Ref.~\cite{datta2020recognizing}, the authors extended this to a more general case, introducing horizon parameters for both objects as $H_1$ and $H_2$. Strictly speaking, these two parameters denote the \textit{fraction} of the flux due to TH in any binary to that in a BBH, and they take values $H_{1,2}\in[0,1]$.

In case of circular orbits, the flux of energy at the horizon can be expressed as a PN expansion~\cite{Alvi:2001mx, Poisson:2018qqd, Poisson:2009di, Nagar:2011aa, Bernuzzi:2012ku,Chatziioannou:2016kem, Cardoso:2012zn}. Since TH implies the presence of horizon, the TH energy flux due to each component has to be multiplied with the corresponding $H_i$.

 Let us consider a compact binary with individual masses $m_1$ and $m_2$, dimensionless spins $\chi_1$ and $\chi_2$, total mass $M=m_1+m_2$ and mass-ratio $q=m_1/m_2$ with $m_1\geq m_2$. In the case of partial absorption, one has $0<H_i<1$. Then the absorbed flux can be expressed as~\cite{datta2020recognizing}
\begin{equation}
\label{dedt}
\begin{aligned}
    -\dv{E}{t} = &{}\frac{32}{5}\nu^2 \frac{v^{15}}{4}\sum_{i=1}^{2} H_i\left(\frac{m^{}_i}{M}\right)^3 \left( 1 + 3\chi^2_i\right)\\&
    \times\left\{-(\hat{L}\cdot\hat{S}_i)\chi^{}_i
     + 2 \left[ 1+\sqrt{1 - \chi^2_i}\right]\frac{m^{}_i}{M}v^3\right \}\ ,
\end{aligned}
\end{equation}
where 
$\nu = {m_1 m_2}/{M^2}$ is the symmetric mass-ratio,  $v$ is the characteristic velocity, and
$\hat{S}^{}_i$ and $\hat{L}$ are the unit vectors along the directions of the $i$th object's spin and the orbital angular momentum, respectively.

There are a few things to note from this expression. This is the expression for the rate of energy \textit{absorption} by the compact object. For a spinless binary, i.e. $\chi^{}_1=\chi^{}_2=0$, the right-hand side survives, meaning that tidal heating is still possible; but in that case it is always positive, which means that the energy flux can only be inward and not outward (which is expected for non-spinning BHs). The presence of the term $-(\hat{L}\cdot\hat{S}_i)\chi^{}_i$ contributes to the loss of energy by the BH, which means that energy is being transferred to the orbit. Also, we see that for anti-aligned spins, where ($\hat{L}\cdot\hat{S}_i$) is negative, energy extraction from the BH is not possible.

The horizon parameters $H_{1,2}$ appear in the GW phase in terms that also include the masses and spins. This makes them {\em degenerate} with those parameters, in that it is more practical to measure the following effective observable parameters instead of $H_{1,2}$:
\begin{subequations}
\label{Eq.Hparams}
\begin{align}
H_{\rm eff5} \equiv &{} \sum_{i=1}^{2}H^{}_i \left(\frac{m^{}_i}{M}\right)^3 \left(\hat{L}\cdot\hat{S}^{}_i\right)\chi^{} _i \left(3 \chi^{}_i{}^2+1\right)\,,\\
H_{\rm eff8} \equiv &{} ~4 \pi  H_{\rm eff5}+\sum^2_{i=1}H^{}_i \left(\frac{m_i}{M}\right)^4 \left(3 \chi^{}_i{}^2+1\right)\nonumber \\
                &\quad\quad\quad\quad\quad\quad\quad \times \left(\sqrt{1-\chi^{}_i{}^2}+1\right)\,.
\end{align}
\end{subequations}
These are analogous to the effective spin parameter $\chi^{}_{eff}$ that was introduced~\cite{Damour:2001tu, Ajith:2011ec} 
for characterizing spinning compact binary waveforms, where a combination of the spin parameters were introduced as a new parameter that can be measured more precisely. The subscripts here denote the fact that $H_{\rm eff5}$ and $H_{\rm eff8}$ appear in the GW phase in 2.5PN and 4PN order, respectively.

If the system is a BBH, as long as any one of the components has a finite spin, both $H_{\rm eff5}$ and $H_{\rm eff8}$ will be nonzero. On the other hand, when both the components of a BBH have vanishing spins, one has $H_{\rm eff5} \to 0$, but  $H_{\rm eff8}\neq 0$. Therefore, in the low-spin limit, $H_{\rm eff8}$ acts as the discriminator for the presence or absence of horizons. A horizonless binary with negligible tidal heating (e.g. binary neutron star) would have both $H_{\rm eff5}$ and $H_{\rm eff8}$ vanish, regardless of their spin values.

Next, we examine the phase contribution in the gravitational waveforms due to TH. This has been calculated in Ref.~\cite{datta2020recognizing} from Refs.~\cite{Tichy:1999pv, Isoyama:2017tbp} to be

\begin{equation}
\begin{aligned}
\label{eq:phase correction1}
\Psi^{}_{\rm TH} = &{} \frac{3}{128\nu} \left(\frac{1}{v}\right)^5 \left[-  \frac{10 }{9 }v^5 H_{\rm eff5} \left(3 \log \left(v\right)+1\right) \right. \\
&-  \frac{5}{168} v^7 H_{\rm eff5} \left(952 \nu +995\right) \\ 
&\left.+   \frac{5}{9}v^8 \left(3 \log \left(v\right)-1\right)(-4 H_{\rm eff8}+ H_{\rm eff5} \Psi^{}_{\text{SO}} )\right]\,,
\end{aligned}
\end{equation}
where the ``spin-orbit" term $\Psi^{}_{\rm SO}$ is given by

\begin{equation}
\label{spin-orbit}
   \begin{split}
        \Psi^{}_{\text{SO}} =&~ \frac{\big(\hat{L}\cdot\hat{S}^{}_1\big)\chi_1 m_1 (73 m_1 + 45 m_2) + 1\leftrightarrow 2}{3 M^2}\,\\
        =&~\frac{73}{3(1+q)^2}\left\{q^2 \big(\hat{L}\cdot\hat{S}^{}_1\big)\chi_1 + \big(\hat{L}\cdot\hat{S}^{}_2\big)\chi_2\right\}\\
        &+ \frac{15q}{(1+q)^2} \left\{ \big(\hat{L}\cdot\hat{S}^{}_1\big)\chi_1 + \big(\hat{L}\cdot\hat{S}^{}_2\big)\chi_2\right\}\,.
   \end{split}
\end{equation}

Equation~\eqref{eq:phase correction1} gives the total phase contribution in the gravitational waveforms due to TH. We can rewrite this expression in a compact form by identifying the dependence of individual terms on $v$ as
\begin{equation}
\begin{aligned}
    \Psi^{}_{\rm TH} = &\, [C_0 + C_1\log(v) + C_2v^2 + C_3v^3\\
    & + C_4v^3\log(v)] H_{\rm eff5} \\
    & + [D_3v^3 + D_4v^3\log(v)]H_{\rm eff8}\,,
\end{aligned}
\end{equation}
where the coefficients $C_i$ ($i=0,1,2,3,4$) and $D_i$ ($i=3,4$) are functions of the symmetric mass-ratio $\nu$, and $C_3,C_4$ also include $\Psi_{\rm SO}$. The term $C_0 H_{\rm eff5}$ is independent of $v$ and thus independent of $f$. Therefore, this term can be absorbed into the phase of coalescence $\phi_c$, which is also independent of $f$. The terms $C_3 v^3 H_{\rm eff5}$ and $D_3 v^3 H_{\rm eff8}$ have $v^3$ dependence, so they are $\propto f$. These terms can be absorbed into the time of coalescence $t_c$, which appears in the total GW phase as $2\pi f t_c$, and is $\propto f$ as well. For these reasons, we discard these three terms from $\Psi^{}_{\rm TH}$, equivalently redefining $\phi_c$ and $t_c$.

We are then left with the terms containing $C_1, C_2, C_4$ and $D_4$; which give us the GW phase due to tidal heating to be
\begin{equation}
\begin{aligned}
\label{eq:phase correction}
\Psi^{}_{\rm TH} = &{} \frac{3}{128\nu}  \left[-  \frac{10 }{3 } H_{\rm eff5} \,  \log \left(v\right) \right. \\
&-  \frac{5}{168} v^2 H_{\rm eff5} \left(952 \nu +995\right) \\ 
&\left.+   \frac{5}{3}v^3  \log \left(v\right)(-4 H_{\rm eff8}+ H_{{eff5}} \Psi^{}_{\text{SO}} )\right],
\end{aligned}
\end{equation}
after putting their expressions from Eq.~\eqref{eq:phase correction1}. We will use this expression for $\Psi^{}_{\rm TH}$ here for our analyses. Throughout the paper, we only consider spins aligned with the orbital angular momentum, so that $\hat{L}\cdot\hat{S}^{}_1=\hat{L}\cdot\hat{S}^{}_2=1$ in Eq.~\eqref{spin-orbit}.

Next, we need the PN approximant to which we will add this phase in order to obtain the complete PN waveform with TH included. For this purpose, we consider the \texttt{TaylorF2} approximant~\cite{tf2PhysRevD.80.084043} upto 3.5PN order ($\sim v^2$), constructed under the ``stationary phase approximation" (SPA)~\cite{cutler_flanagan_PhysRevD.49.2658}. Since PN expansions fail near the merger phase due to violations of the slow motion and weak gravity conditions, we have to truncate the waveform at some point where the binary is still away from the merger. A general choice for such cut-off frequency is the binary's {\it innermost stable circular orbit} (ISCO), which marks the ``end" of the inspiral phase. For a binary of KBHs, location of the ISCO depends on the spin-alignment as well as the component masses and spins. In the case of aligned spins, the ISCO for a KBH is closer to the center of mass of the binary than a Schwarzschild BH of the same mass. In our work, we consider the upper cutoff frequency to be the GW frequency at the ISCO corresponding to the final BH formed after merger, given by (ignoring cosmological redshift)~\cite{Favata:2021vhw} 
\begin{equation}
\label{isco}
    f_{\rm \tiny{ISCO}}=\frac{\hat{\Omega}_{\rm \tiny{ISCO}}(\chi^{}_f)}{\pi M_f}\,.
\end{equation}
Here $\hat{\Omega}_{\rm \tiny{ISCO}}(\chi)=M_{\rm \tiny{Kerr}}\Omega_{\rm \tiny{ISCO}}$ is the dimensionless angular frequency for a circular equatorial orbit around a KBH with mass $M_{\rm \tiny{Kerr}}$ and spin $\chi$~\cite{1972ApJ...178..347B}. For the upper cutoff frequency, we choose $M_{\rm \tiny{Kerr}}=M_f$ and $\chi=\chi^{}_f$, the final mass and spin of the merger remnant BH, which are obtained by using fitting formulas from NR simulations~\cite{Husa:2015iqa}. Explicit expressions for $\hat{\Omega}_{\rm \tiny{ISCO}}$, and $M_f,\chi^{}_f$ in terms of initial masses and spins are mentioned in Appendix C of Ref.~\cite{Favata:2021vhw}.


\section{basics of the fisher matrix approach}
\label{fisher_basics}

In this work, we mainly focus on Fisher matrix analysis~\cite{Vallisneri:2007ev, cutler_flanagan_PhysRevD.49.2658, owen_PhysRevD.53.6749, poisson_will_PhysRevD.52.848} for the estimation of errors in the measurement of the horizon parameters in the 3rd generation detectors Einstein Telescope~\cite{maggiore2020science} and Cosmic Explorer~\cite{regimbau2017digging, vitale2017parameter, Reitze:2019iox}. In this section we will briefly summarize the basic concepts of the Fisher matrix approach for parameter estimation. 

A GW signal in the time domain, as emitted by a coalescing compact binary, can be decomposed into two polarization states $h_+(t;\Theta_{\rm GW})$ and $h_{\cross}(t;\Theta_{\rm GW})$, where the parameter vector $\Theta_{\rm GW}$ contains information about the source. For a BBH in PP approximation, $\Theta_{\rm GW}\equiv\{m_1,m_2,\boldsymbol{\chi^{}_1},\boldsymbol{\chi^{}_2},D^{}_L,\iota,t_c,\phi_c\}$, where $m_1,m_2$ are companion masses, $\boldsymbol{\chi^{}_1},\boldsymbol{\chi^{}_2}$ are their dimensionless spin vectors, $D^{}_L$ is the luminosity distance of the binary, $\iota$ is the inclination angle of its orbital plane with respect to the line of sight, and $t_c$ and $\phi_c$ are the time and phase of coalescence, respectively.  We extend this set by including the two parameters $\{H_{\rm eff5}, H_{\rm eff8}\}$, defined in Eq.~\eqref{Eq.Hparams}, to incorporate TH. The GW strain in the frequency domain as measured by a detector, $H(f)$, depends on $\Theta_{\rm GW}$, the location of the detector, and three more extrinsic source parameters $\{\alpha, \delta, \psi\}$, which denote right ascension, declination and polarization angle, respectively.

\subsection{The Noise Power Spectral Density} The ability of a GW detector to measure the GW strain depends on its sensitivity, which in turn depends on the \textit{power spectral density} (PSD) of its noise, $n(t)$, and its auto-correlation~\cite{Borhanian:2020ypi} $\kappa=\overline{n(t_1)n(t_2)}$, where the overbar denotes an average over noise realizations. Assuming that the noise is stationary and Gaussian with zero mean, which means $\kappa$ only depends on the time difference $t' =t_1-t_2$, the PSD of the noise (in frequency domain) can be written as,
\begin{equation}
    S_n(f)=\frac{1}{2}\int_{-\infty}^\infty \dd t' \kappa(t')e^{i2\pi ft'}, \quad \text{with}\quad f>0.
\end{equation}
This function denotes the detector sensitivity at different frequencies.

\subsection{The Signal-to-Noise Ratio} The set of all possible detector responses in frequency or time domain forms a vector space. In frequency domain, let us call this space $\mathcal{V}$. We can define, on this space, a \textit{noise-weighted scalar product} of two detector responses $H(f), G(f) \in \mathcal{V}$ as~\cite{Sathyaprakash:2009xs}
\begin{equation}
\label{innerproduct}
   \braket{H}{G}=2\int^\infty_0 \dd f \frac{H^\ast(f) G(f) + G^\ast (f)H(f)}{S_n(f)}.
\end{equation}
Equipped with this definition, we can define the {\it signal-to-noise ratio} (SNR) $\rho$ for a given GW signal $H$ as
\begin{equation}
\label{snr}
    \rho=\sqrt{\braket{H}}=2\sqrt{\int^\infty_0 \dd f \frac{|H(f)|^2}{S_n(f)}},
\end{equation}
where $S_n(f)$ contains information about the sensitivity of the chosen detector. The SNR of a signal characterizes its loudness over a given noise profile.

It is important to mention here that in practical situations, like in this study, we will not cover the entire frequency region (0 to $\infty$), because detectors typically have sensitivity only within a finite frequency band, and the signal band is also finite. So, Eq.~\eqref{snr} will be replaced by
\begin{equation}
\label{snr2}
    \rho=2\sqrt{\int^{f_{\rm max}}_{f_{\rm min}} \dd f \frac{|H(f)|^2}{S_n(f)}}.
\end{equation}
Here $f_{\rm min}$ will be determined by the detector band's lower frequency cut-off, and $f_{\rm max}$ will correspond to the ISCO (Eq.~\eqref{isco}).

\subsection{The Fisher Information Matrix}
\label{fisher_matrix}
The detector output $S(f)$ in frequency domain is related to the GW strain $H(f)$ and the noise $N(f)$ as $S(f)=H(f)+N(f)$. Since we have assumed a Gaussian profile for the noise, we can write the probability function for $N(f)$~\cite{Borhanian:2020ypi} as
\begin{align}
\label{ptheta}
     p(\Theta)=&p_0(\Theta)e^{-\frac{1}{2}\braket{N}}\nonumber\\
     =&p_0(\Theta)e^{-\frac{1}{2}\braket{S-H(\Theta)}},
\end{align}
where $\Theta$ is the parameter vector and $p_0$ is the prior on these parameters.

Let us denote $E(\Theta)=\braket{S-H(\Theta)}$, and expand this quantity around the ``true" value ($\Theta^\ast$) of the parameters :
\begin{equation}
\label{etheta}
    E(\Theta)=E(\Theta^\ast)+\frac{1}{2}\pdv{E}{\Theta_i}{\Theta_j}\bigg|_{\Theta=\Theta^\ast}\Delta\Theta^i\Delta\Theta^j+\cdots,
\end{equation}
where $\Delta\Theta = (\Theta-\Theta^\ast)$, and we use Einstein summation convention over repeated indices. Also, using the expression for $E(\Theta)$, we can write
\begin{align}
    \pdv{E(\Theta)}{\Theta_i}{\Theta_j} = &~ 2\braket{\partial_{\Theta_i}H(\Theta)}{\partial_{\Theta_j}H(\Theta)}+\braket{\partial_{\Theta_i}\partial_{\Theta_j}H(\Theta)}{N}\nonumber\\
    \approx &~ 2\braket{\partial_{\Theta_i}H(\Theta)}{\partial_{\Theta_j}H(\Theta)},
\label{dedtheta}
\end{align}
where in the second step we have assumed that the SNR value is high enough for the first order derivatives of $H$ to dominate over the second order ones~\cite{cutler_flanagan_PhysRevD.49.2658}.

We now define the {\it Fisher information matrix} $\Gamma$, the elements of which are given as
\begin{equation}
\label{gamma}
    \Gamma_{ij}=\braket{\partial_{\Theta_i}H(\Theta)}{\partial_{\Theta_j}H(\Theta)}.
\end{equation}
Using this in Eq.~\eqref{dedtheta}, and assuming that $\Delta\Theta$ is small, we can infer from Eq.~\eqref{ptheta} that
\begin{equation}
    p(\Theta)\propto\exp{-\frac{1}{2}\Gamma_{ij}\Delta\Theta^i\Delta\Theta^j}.
\end{equation}
The inverse of the Fisher matrix is the \textit{covariance matrix}, $C=\Gamma^{-1}$. Along the diagonal of $C$, one gets the variances of the concerned parameters, from which one can get the 1-$\sigma$ errors in those parameters as $\sigma_{\Theta_i}=\sqrt{C_{ii}}$. The off-diagonal elements are the covariances between the parameters, defined as
\begin{equation}
    C_{ij}={\rm cov}(\Theta_i,\Theta_j)=\overline{(\Theta_i-\overline{\Theta}_i)(\Theta_j-\overline{\Theta}_j)},
\end{equation}
where the bar denotes mean value. For $i=j$, one gets $C_{ii}=\overline{(\Theta_i-\overline{\Theta}_i)^2}$, called the ``variance" of the distribution of $\Theta_i$, which is the square of its standard deviation $\sigma_{\Theta_i}$.

For the Fisher matrix approach to work, not only the SNR has to be high, but the matrix also has to be {\it well-conditioned}~\cite{Borhanian:2020ypi}. This criterion is quantified by the {\it condition number}, which is defined as the ratio of the largest and the smallest eigenvalues of the matrix. If this quantity is too large, then the inversion of $\Gamma$ is not trustworthy. Here, we ensured that it is well within the numerical precision available for our computations~\cite{condnum_PhysRevD.85.062002}.



\begin{figure}[]
\centering     
\subfigure[~Errors in $H_{\rm eff5}$ in LIGO-Virgo]{\label{h5-ligo-m}\includegraphics[width=85mm]{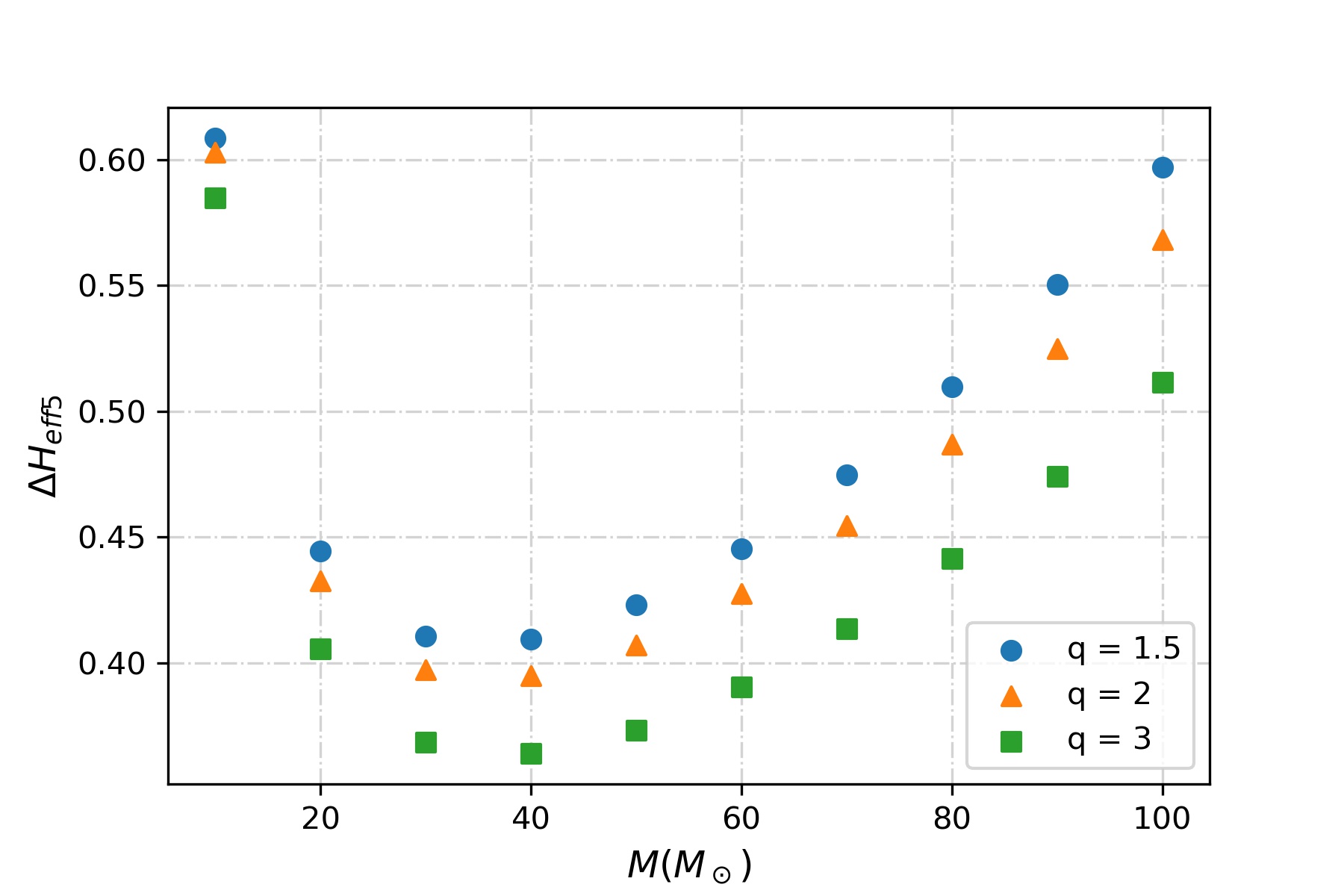}}
\subfigure[~Errors in $H_{\rm eff8}$ in LIGO-Virgo]{\label{h8-ligo-m}\includegraphics[width=85mm]{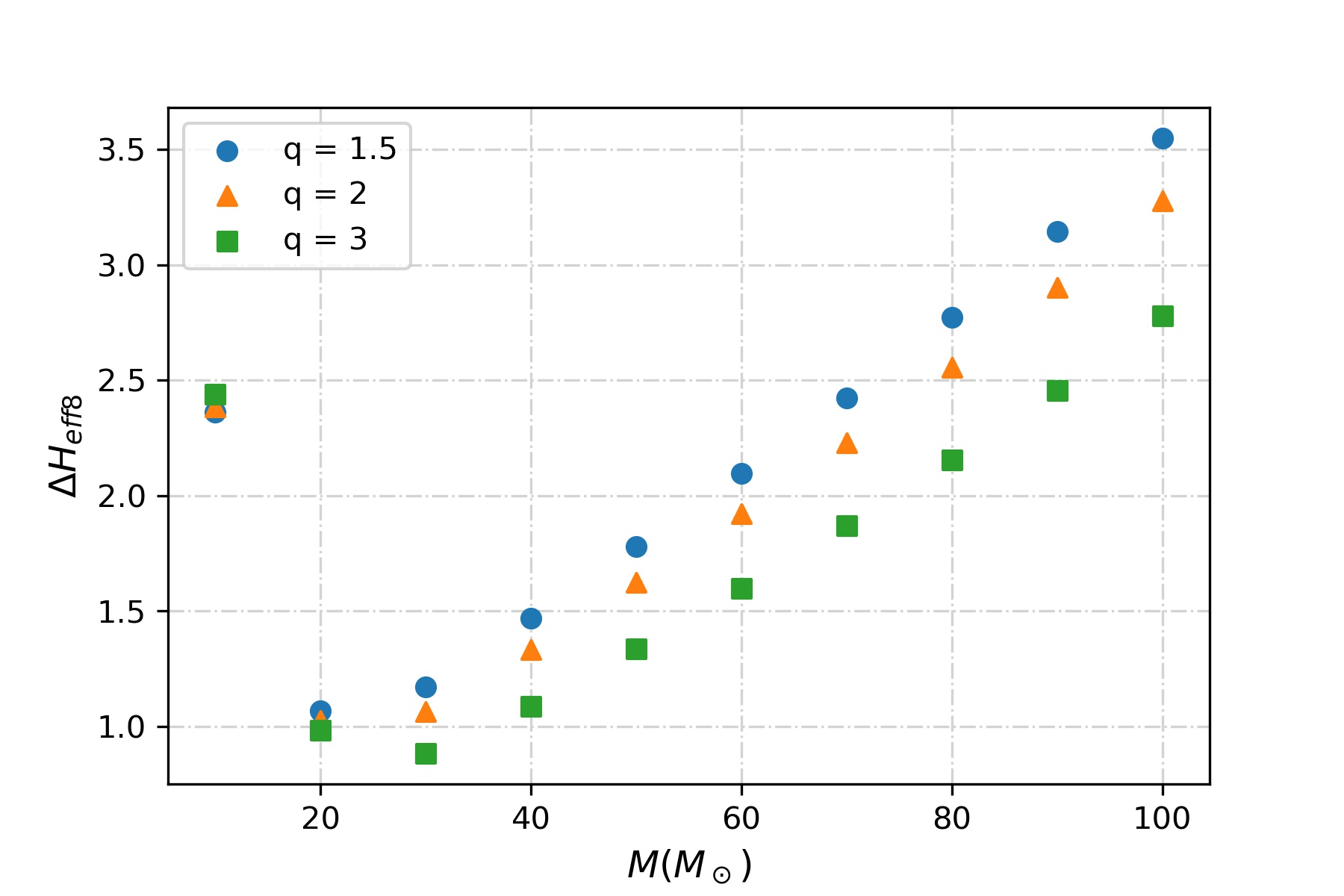}}
\caption{\small{Error values in the TH parameters $H_{\rm eff5}$ (top) and $H_{\rm eff8}$ (bottom) as a function of total mass, when measured by the three detector network of LIGO (Hanford, Livingston) and Virgo.  mass-ratio ($q$) has been varied from 1.5 to 3 for getting different curves. We consider aligned spins here, so that $\hat{L}\cdot\hat{S}_i=1$. $H_{\rm eff5}=0.6, H_{\rm eff8}=12, D^{}_L=200$ Mpc, $\chi^{}_1=\chi^{}_2=0.8$ have been taken.}} 
\label{ligo-m}
\end{figure}


\begin{figure*}[ht]
\centering     

\subfigure[~Errors in $H_{\rm eff5}$ in Einstein Telescope]{\label{h5-et-m}\includegraphics[width=85mm]{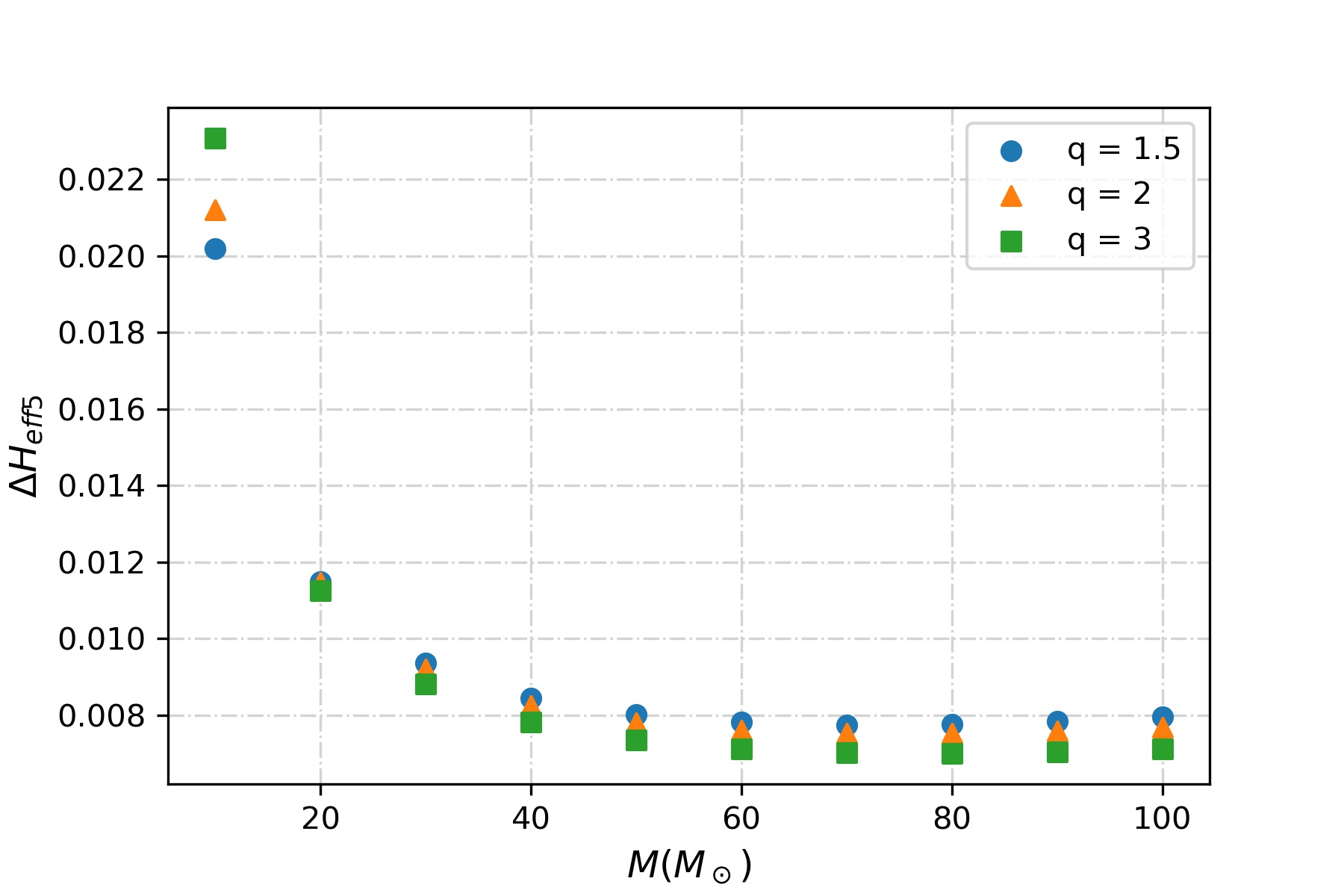}}
\subfigure[~Errors in $H_{\rm eff5}$ in Cosmic Explorer]{\label{h5-ce-m}\includegraphics[width=85mm]{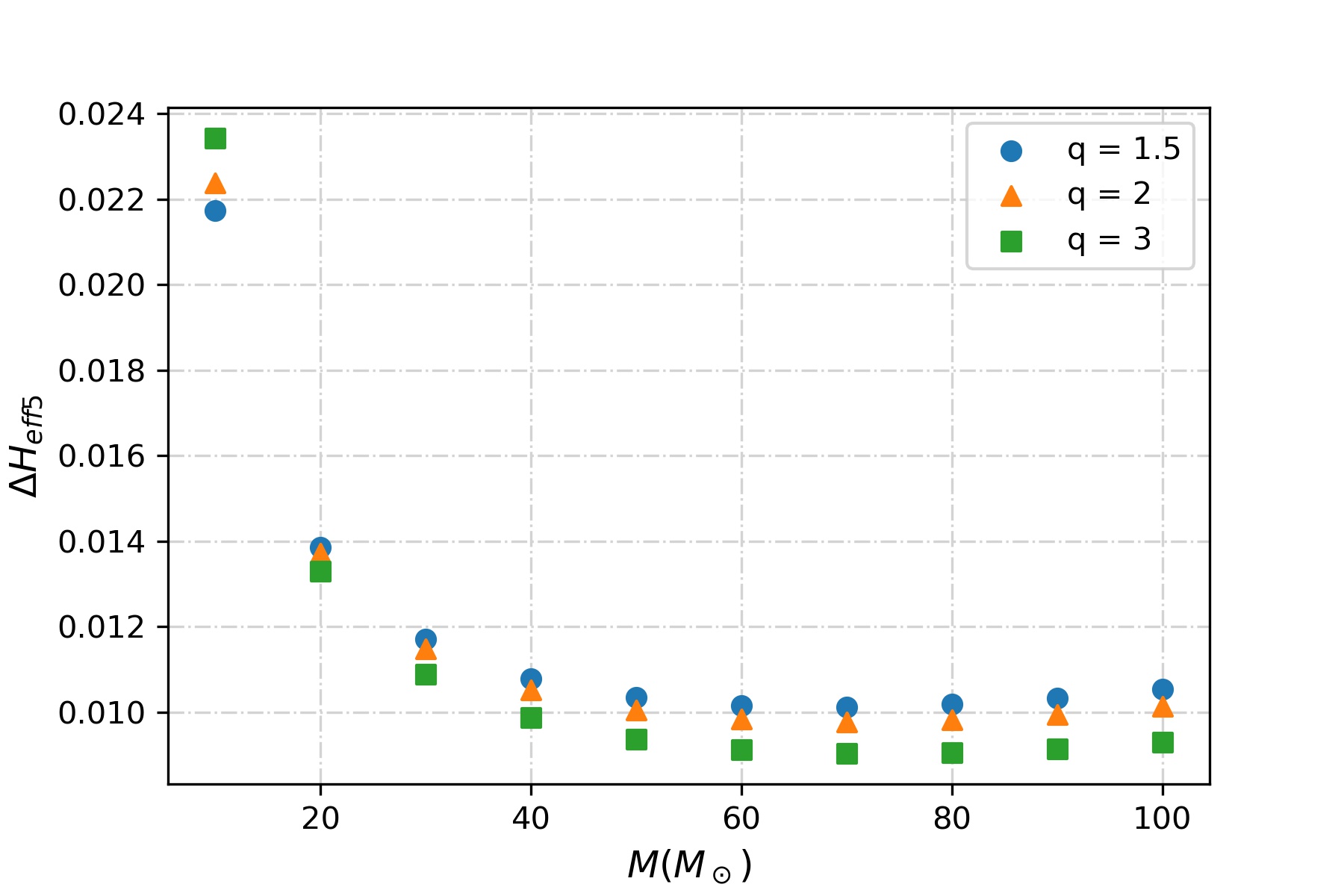}}
\subfigure[~Errors in $H_{\rm eff8}$ in Einstein Telescope]{\label{h8-et-m}\includegraphics[width=85mm]{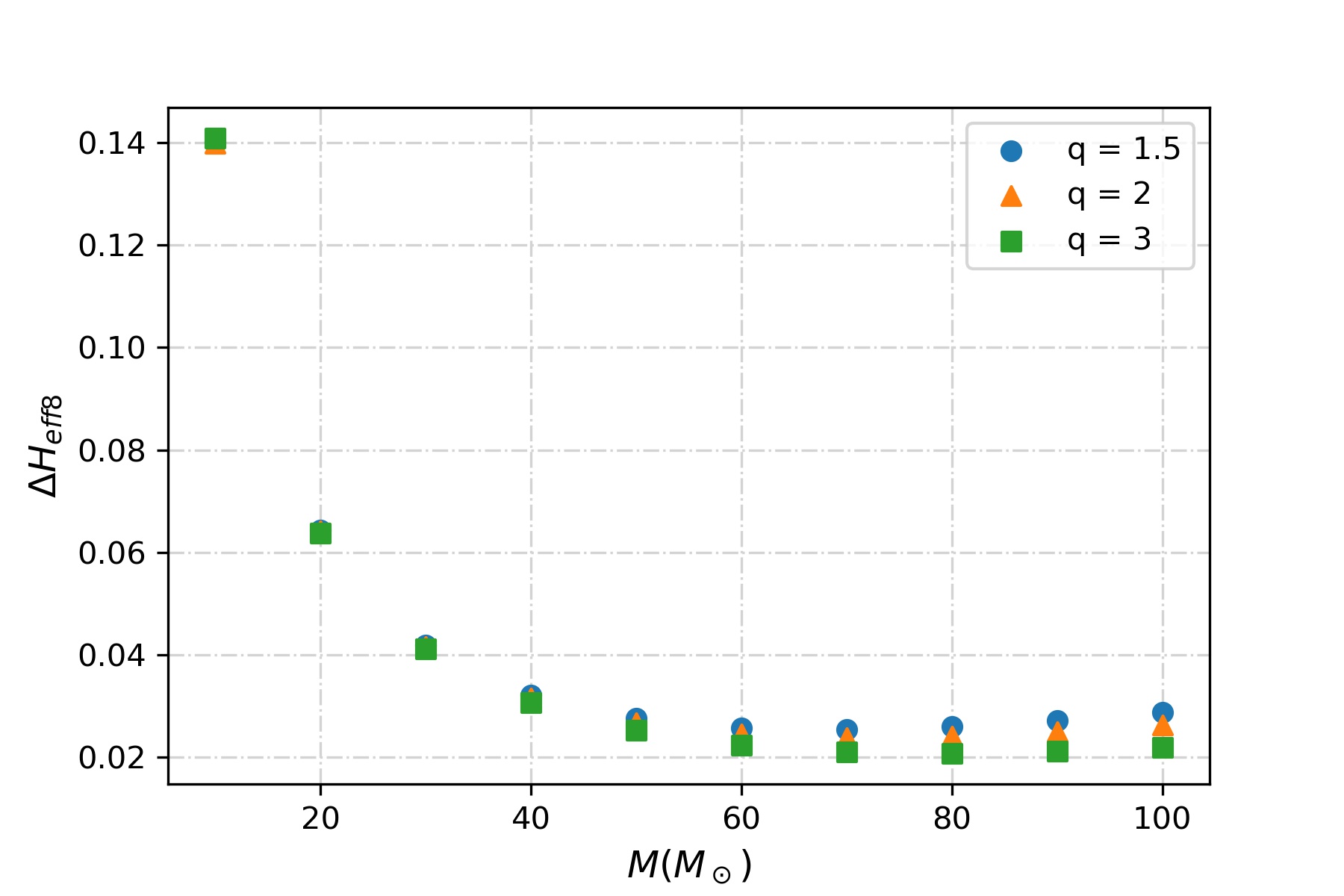}}
\subfigure[~Errors in $H_{\rm eff8}$ in Cosmic Explorer]{\label{h8-ce-m}\includegraphics[width=85mm]{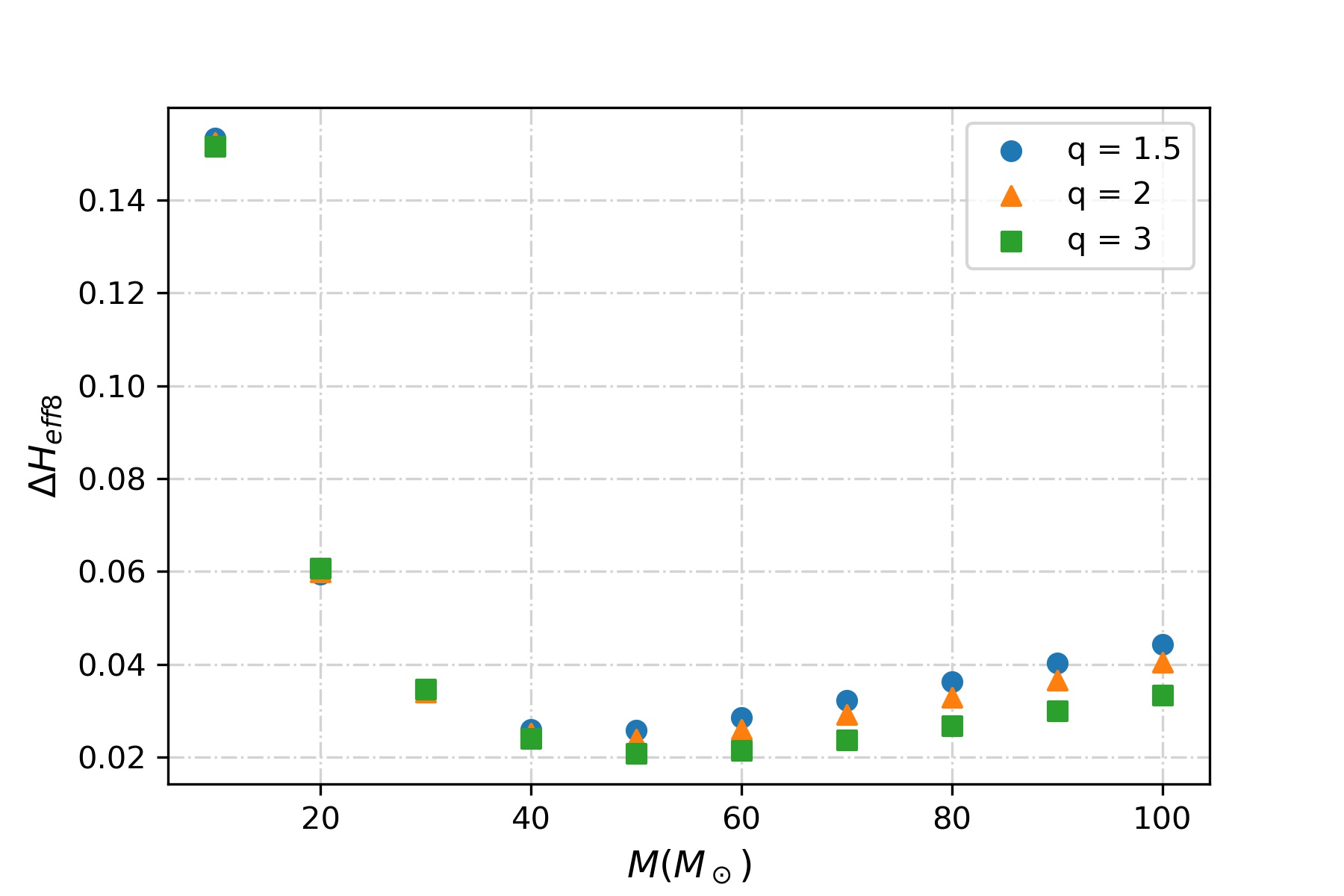}}

\caption{\small{Errors in $H_{\rm eff5}$ (top row) and $H_{\rm eff8}$ (bottom row) as a function of total mass $M$, when measured in ET (first column) and CE (second column). Injection parameters are the same as in Fig.~\ref{ligo-m}. }}
\label{et-ce-m}
\end{figure*}


\begin{figure*}[ht]

\centering     

\subfigure[~Errors in $H_{\rm eff5}$ in Einstein Telescope]{\label{h5-et-dl}\includegraphics[width=8.5cm]{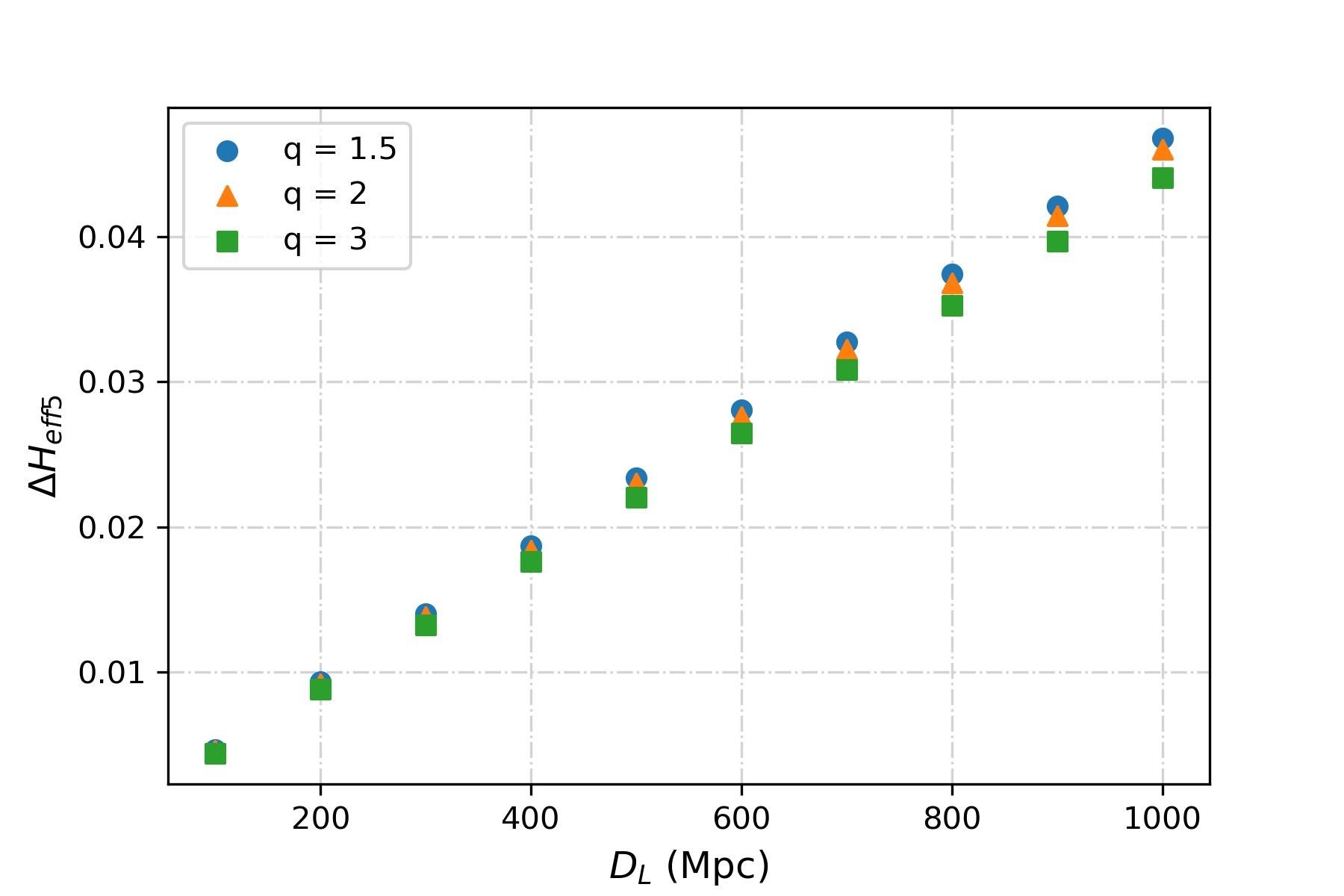}}
\subfigure[~Errors in $H_{\rm eff5}$ in Cosmic Explorer]{\label{h5-ce-dl}\includegraphics[width=8.5cm]{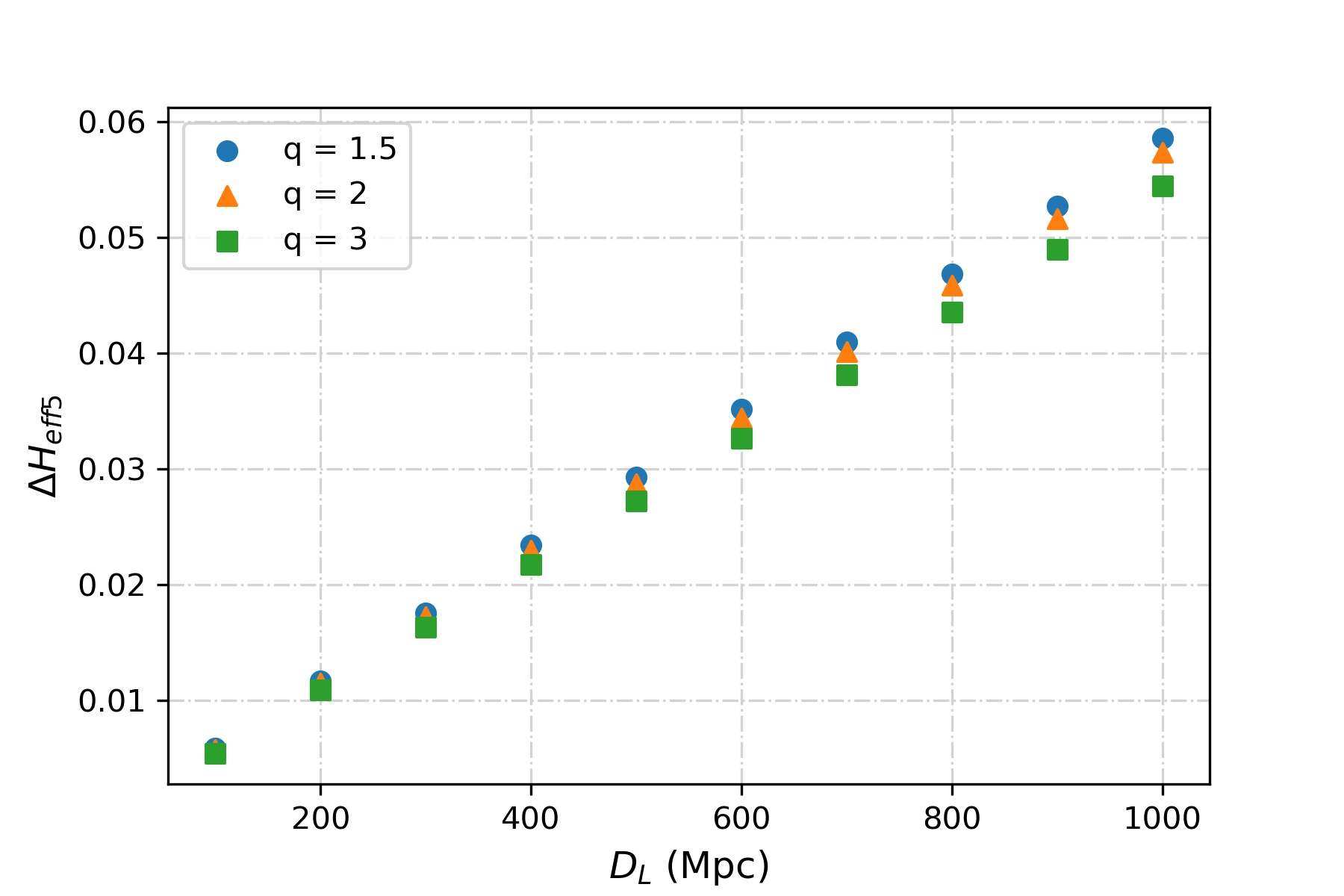}}
\subfigure[~Errors in $H_{\rm eff8}$ in Einstein Telescope]{\label{h8-et-dl}\includegraphics[width=8.5cm]{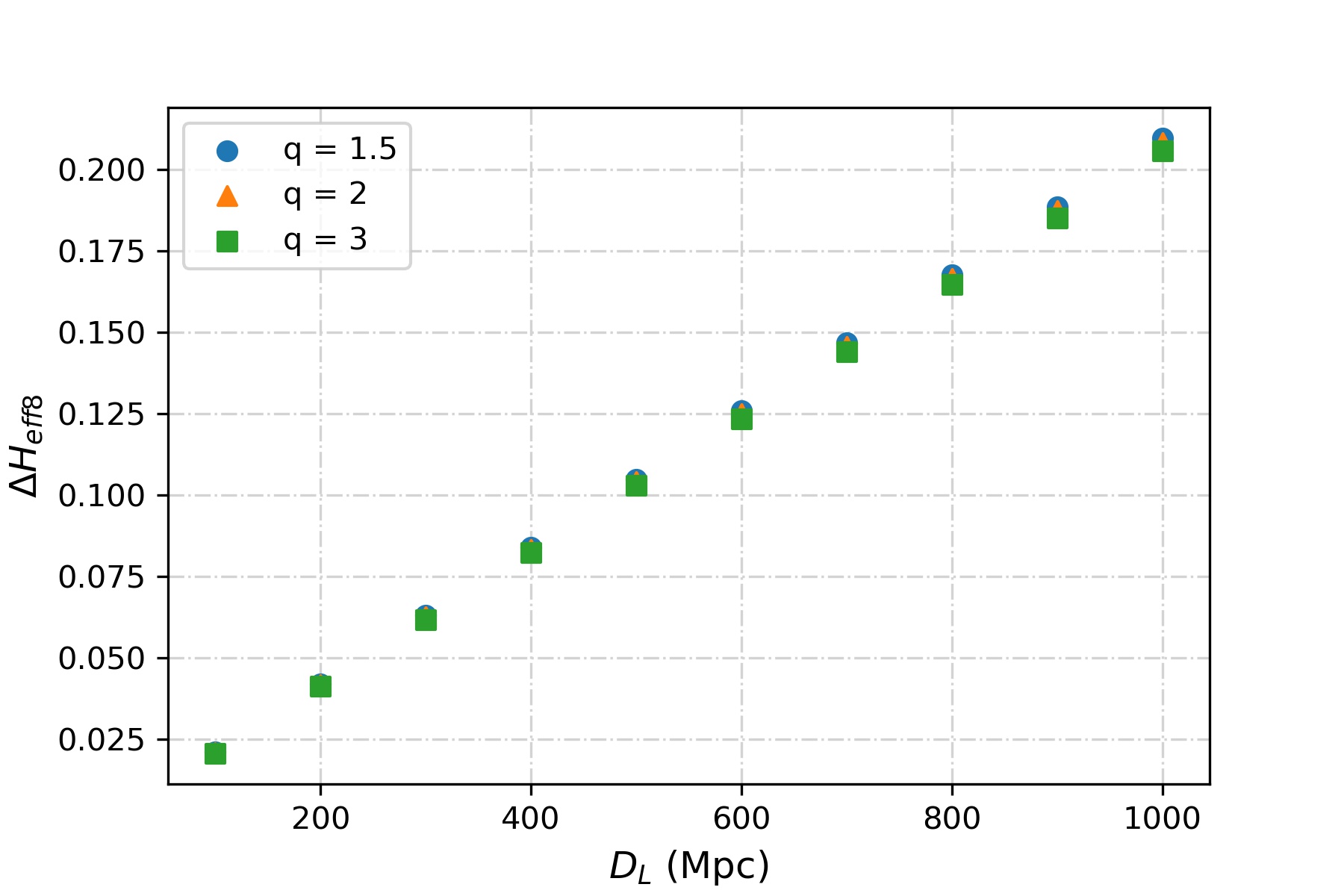}}
\subfigure[~Errors in $H_{\rm eff8}$ in Cosmic Explorer]{\label{h8-ce-dl}\includegraphics[width=8.5cm]{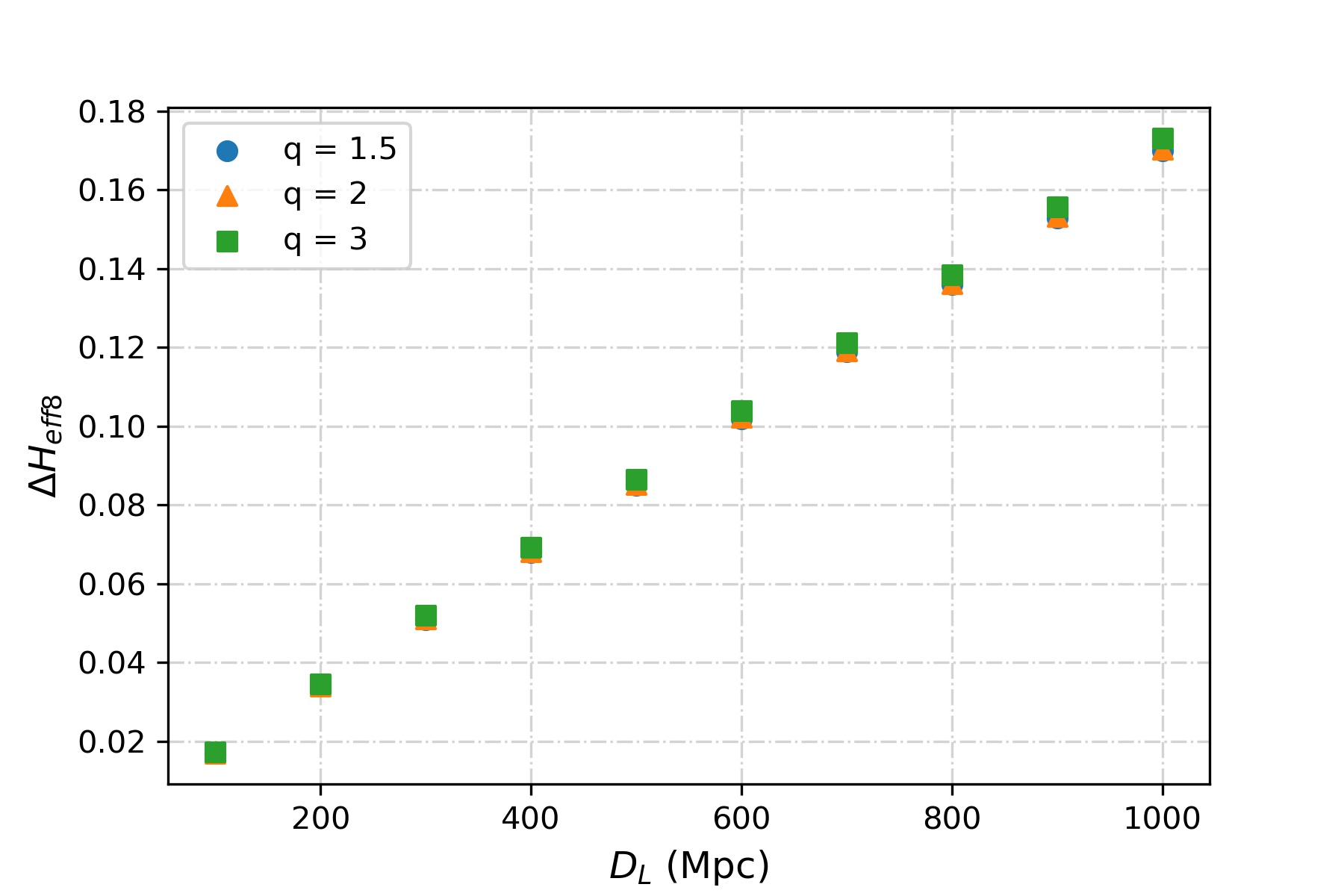}}

\caption{\small{Errors in $H_{\rm eff5}$ (top row) and $H_{\rm eff8}$ (bottom row) as a function of luminosity distance, when measured in ET (first column) and CE (second column). Along the X-axis, $D^{}_L$ varies from 100 Mpc to 1 Gpc. Other parameters are fixed at $H_{\rm eff5}=0.6, H_{\rm eff8}=12, M=30M_\odot$, $\chi^{}_1=\chi^{}_2=0.8$.}}

\label{et-ce-dl}
\end{figure*}


\section{Results of fisher analyses}
\label{results}

In this section we will apply the Fisher matrix approach to estimate the errors in the TH parameters in the three-detector network comprising the Advanced LIGO~\cite{LIGOScientific:2014pky} and Advanced Virgo~\cite{VIRGO:2014yos} detectors, and the proposed 3rd generation detectors Einstein Telescope~\cite{maggiore2020science} and Cosmic Explorer~\cite{Reitze:2019iox}. We have used the package \texttt{GWBENCH}~\cite{Borhanian:2020ypi} for our Fisher matrix calculations. In order to estimate the errors in certain parameters using this approach, we first inject a gravitational waveform into the relevant detector, then using Eq.~\eqref{gamma}, calculate the Fisher matrix $\Gamma$ and the covariance matrix by inverting it. As mentioned in Sec.~\ref{theory_TH}, in this work we take \texttt{TaylorF2} as the PN approximant and incorporate in it the phase contribution ($\Psi^{}_{\rm TH}$) due to TH, given by Eq.~\eqref{eq:phase correction}. 

In our study, we estimate the projected errors in five parameters, $\Theta \equiv\{\mathcal{M}_c,\nu,D_L,H_{\rm eff5},H_{\rm eff8}\}$, where $\mathcal{M}_c$ is the \textit{chirp mass} defined as $\mathcal{M}_c=(m_1m_2)^{3/5}/M^{1/5}$. When we discuss the variation of the errors with component spins in Sec.~\ref{sec:spin}, we extend the parameter space with the two component spins $\chi_1,\chi_2$. As the lower cutoff frequencies, we have used 10 Hz (4 Hz) for LIGO-Virgo (ET, CE), and the upper cutoff frequencies are determined by the spin-dependent ISCO frequencies given by Eq.~\eqref{isco}.

From Eqs.~\eqref{Eq.Hparams} we see that $H_{\rm eff5}$ and $H_{\rm eff8}$ are functions of the component masses only through the ratios $m_1/M=q/(1+q)$ and $m_2/M=1/(1+q)$, both of which have values always lying between 0 and 1. Also, $-1\leq\chi^{}_i\leq 1$. This enables one to define a range in the values of these two parameters that can occur physically, for all possible values of $q$. This turns out to be approximately~\cite{datta2020recognizing}
\begin{equation}
\label{range}
    -4\leq H_{\rm eff5}\leq 4\,,  \quad {\rm and} \quad -46.3\lesssim H_{\rm eff8}\lesssim 54.3\,.
\end{equation}
Even though we will treat $H_{\rm eff5}$ and $H_{\rm eff8}$ as free parameters here, we have to keep in mind that this is the physical range of values they can have.

\subsection{LIGO \& Virgo}
\label{ligo-results}

Figure~\ref{ligo-m} shows the variation of 1-$\sigma$ errors with the total binary mass in the noise spectrum of the three-detector network of LIGO (Hanford, Livingston) and Virgo. The Y-axes report the 1-$\sigma$ errors in $H_{\rm eff5}$ (Fig.~\ref{h5-ligo-m}) and $H_{\rm eff8}$ (Fig.~\ref{h8-ligo-m}), denoted by $\Delta H_{\rm eff5}$ and $\Delta H_{\rm eff8}$, respectively. In our analysis, we have used the most recent design sensitivity curves of Advanced LIGO~\cite{LIGO:design-sensitivity} and Advanced Virgo~\cite{Virgo:design-sensitivity} detectors. Binaries in the range of total mass $10-100M_\odot$ have been considered, situated at a distance of 200 Mpc. The errors initially fall with total binary mass in the range $M<30M_\odot$, and we find that there is a region around $30 M_\odot$ where the errors are minimum. Thereafter, the errors rise rapidly with increasing $M$. An increase in $M$ causes the SNR to rise, which provides a better estimation for $H_{\rm eff5}, H_{\rm eff8}$. This causes the dip in the errors for $M = 10-30 M_\odot$. Further increase in $M$ shrinks the signal band. This is because it lowers the value of the ISCO  frequency while $f_{\rm min}$ remains fixed. This interplay between the SNR and the effective frequency interval creates an optimal region, which turns out to be about $30-40 M_\odot$. We also note that the exact minima in the errors are slightly different for $H_{\rm eff5}$ ($\sim 40 M_\odot$) and $H_{\rm eff8}$ ($\sim 30 M_\odot$).

$H_{\rm eff5}$ and $H_{\rm eff8}$ for binaries with more asymmetric masses appear to be more precisely measurable. This is expected because mass asymmetry lowers the value of the symmetric mass-ratio $\nu$, and the TH phase has a prefactor of $1/\nu$ (see Eq.~\eqref{eq:phase correction}), making the phase contribution due to TH higher for more asymmetric masses, consequently adding more GW cycles into the signal band.  For a binary at 200 Mpc with $M=30M_\odot$ and $q=1.5$, Fisher estimates show an error value of $\sim 0.4 (1.2)$ for $H_{\rm eff5} (H_{\rm eff8})$, which amounts to a relative percentage error of $\sim 67\%(10\%)$. In LIGO, then, the detection of a so-called {\it golden binary} at a distance $\leq 50$ Mpc will make it possible to estimate these TH parameters with better than 17\% precision.

\subsection{3rd Generation Detectors}
\label{3g-results}

The proposed 3rd generation (3G) GW detectors, Einstein Telescope (ET) and Cosmic Explorer (CE) will have a higher sensitivity than current detectors, which will result in higher SNR for CBCs. This makes Fisher error projections quite trustworthy. In this section, we explore the measurement precision of $H_{\rm eff5},H_{\rm eff8}$ in ET and CE. In addition to the variations of the errors with $M$, we will also look at the variations with luminosity distance $D^{}_L$ and the spin values $\chi^{}_{1,2}$.

For our study, we have used the sensitivity curves for the ET-D configuration \cite{Hild:2010id} of Einstein Telescope, and 40 km long CE configuration \cite{CE:sensitivity, LIGOScientific:2016wof} of Cosmic Explorer, optimized for the low fequencies of CBC.

\subsubsection{Dependence on the Total Binary Mass}


\begin{figure}[]
    \centering
    \includegraphics[width=85mm]{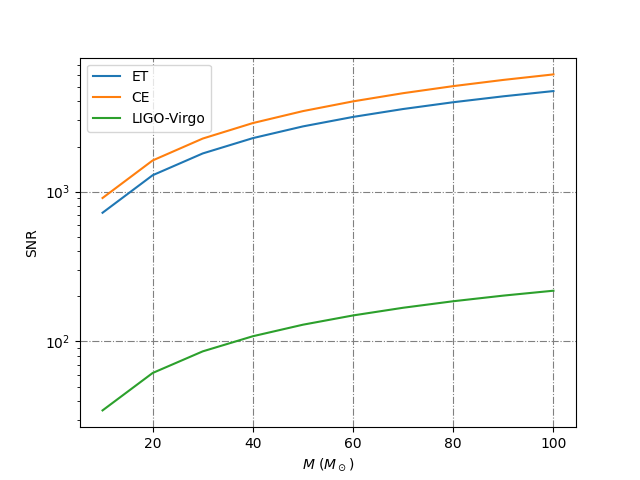}
    \caption{\small Variation of SNR with the total mass $M$ in LIGO-Virgo, ET and CE, as calculated from Eq.~\eqref{snr}. We consider binaries at $D_L = 200$ Mpc with mass-ratio $q=1.5$. }
    \label{snr-m}
\end{figure}

Figure~\ref{et-ce-m} shows the variation of errors with total mass.  As expected, the errors are smaller compared to LIGO-Virgo, due to the high SNR values. In 3G detectors (Fig.~\ref{snr-m}), typically SNR $\sim\mathcal{O}(10^3)$ whereas in LIGO-Virgo, SNR $\sim\mathcal{O}(10^2)$ for the chosen parameter space. Also, SNR increases more rapidly with $M$ in ET and CE than in LIGO-Virgo,  making the rise in errors due to the shortening of the frequency range much slower after $M\sim 60M_\odot$, as seen from Fig.~\ref{et-ce-m}. Comparing Fig.~\ref{ligo-m} and Fig.~\ref{et-ce-m}, we confirm that the precision of measurement in 3G detectors has substantial improvement over LIGO-Virgo.

 For binaries with $M\gtrsim 60M_\odot$ and $q=1.5$, estimation of $H_{\rm eff5}$ and $H_{\rm eff8}$ can be made with 1-$\sigma$ errors $\Delta H_{\rm eff5}\sim$ 0.008 (1.2\%) and $\Delta H_{\rm eff8} \sim 0.02$  (0.22\%) respectively, for BBHs at a distance of 200 Mpc. The error values fall as more component mass asymmetry is introduced. However, for binaries with low masses, we see that this trend is reversed for $H_{\rm eff5}$ -- as seen in Figs.~\ref{h5-et-m} and~\ref{h5-ce-m} for $M=10 M_\odot$. We also note that the variation in the error values with changing $q$ is less pronounced in ET, CE than  LIGO-Virgo.


\begin{figure}[]
    \centering
    \includegraphics[width=86mm]{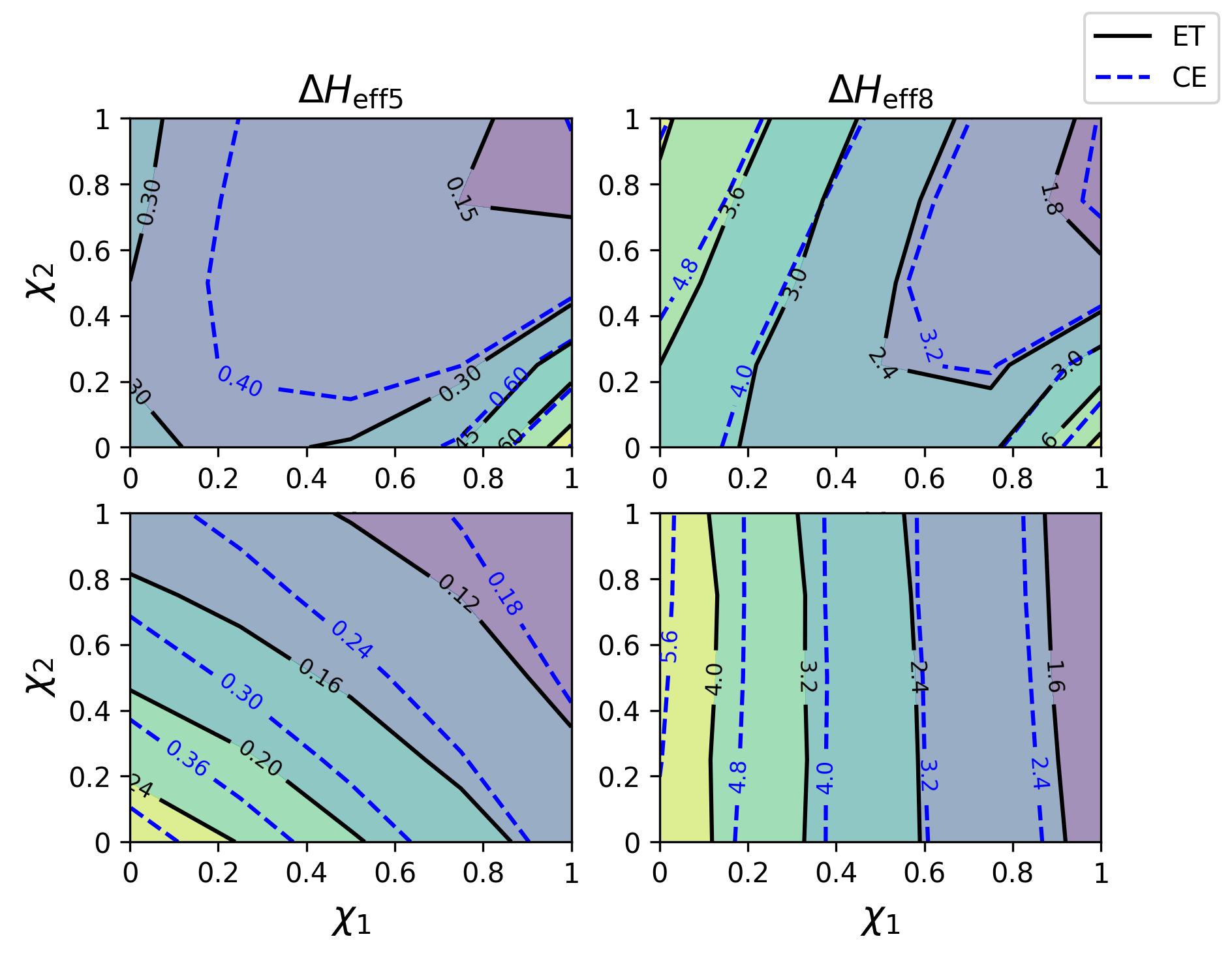}
    \caption{\small Variation of errors in $H_{\rm eff5}$ (left column) and $H_{\rm eff8}$ (right column) in ET (solid lines), CE (dashed lines) with dimensionless spins. $\chi_1, \chi_2$ are varied from 0 to 1 along the $X$ and $Y$-axes respectively. Total binary mass is $M=40 M_\odot$, and mass-ratios are $q=1.1$ (top panel) and $q=3$ (bottom panel). We consider optimally oriented binaries at $D_L = 200$ Mpc, with $H_{\rm eff5}=0.6$, and $H_{\rm eff8}=12$.}
\label{spinvar}
\end{figure}

\subsubsection{Dependence on the Luminosity Distance}
\label{sec:distance}

Figure~\ref{et-ce-dl} shows the variation of errors with luminosity distance $D^{}_L$. We keep the total mass fixed at $M=30M_\odot$. In this case only the fall in SNR with increasing distance affects the errors. As expected, the errors rise linearly with $D^{}_L$, with the slope being greater for more symmetric masses. From Fig.~\ref{et-ce-dl}, we see that for binaries as far as 1 Gpc away, the 1-$\sigma$ error in $H_{\rm eff5}$ is $\sim$ 0.06 (10\%) for BH binaries with $q=3$, whereas for $H_{\rm eff8}$ it is $\sim$ 0.2 (1.6\%). Owing to the linear variation in errors, the presence of horizons for all the sources within this range can be tested with linearly increasing precision.


\begin{figure*}[ht]
\centering     
\subfigure[~LIGO-Virgo]{\label{40-ligo}\includegraphics[width=0.48\textwidth]{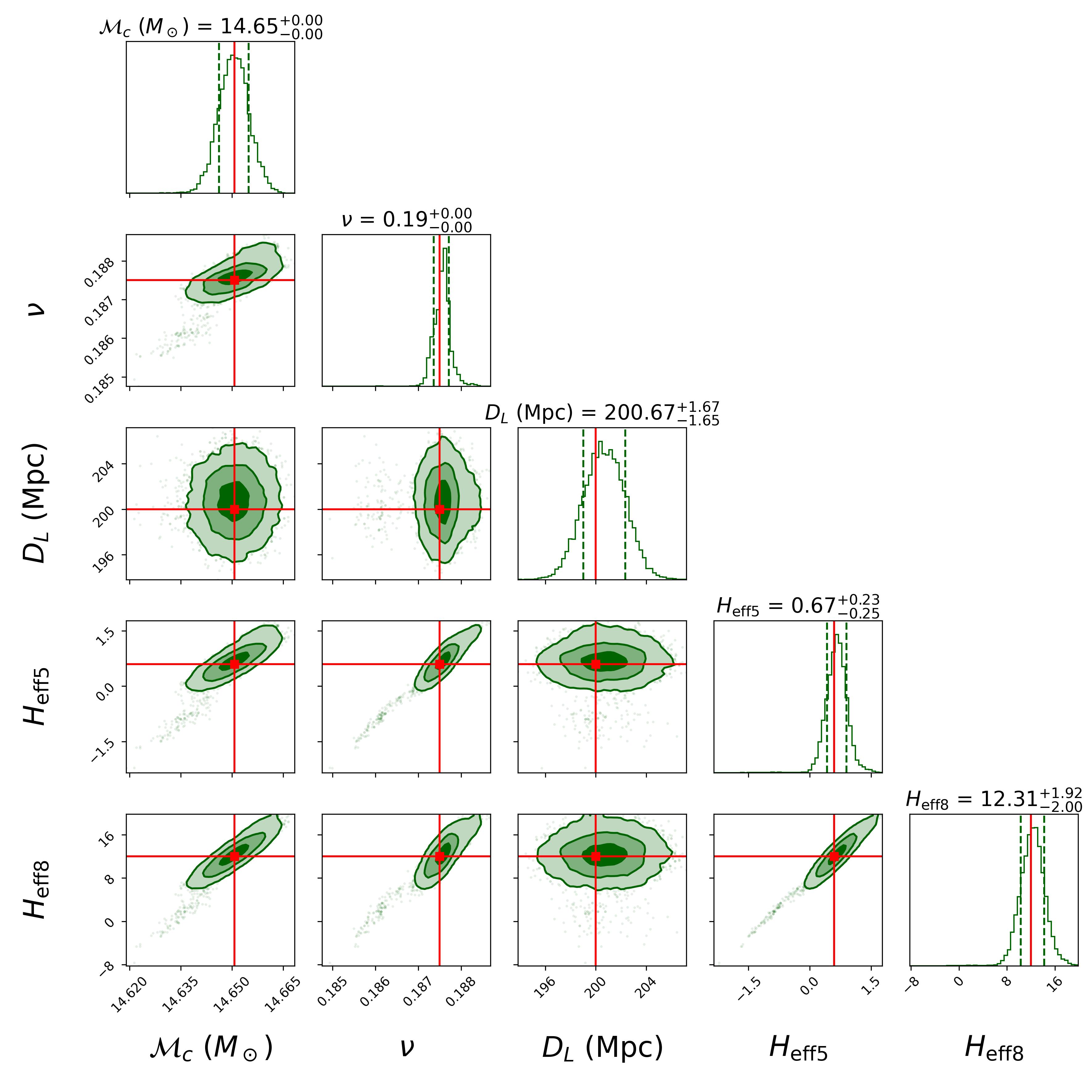}}
\subfigure[~Einstein Telescope]{\label{40-et}\includegraphics[width=0.48\textwidth]{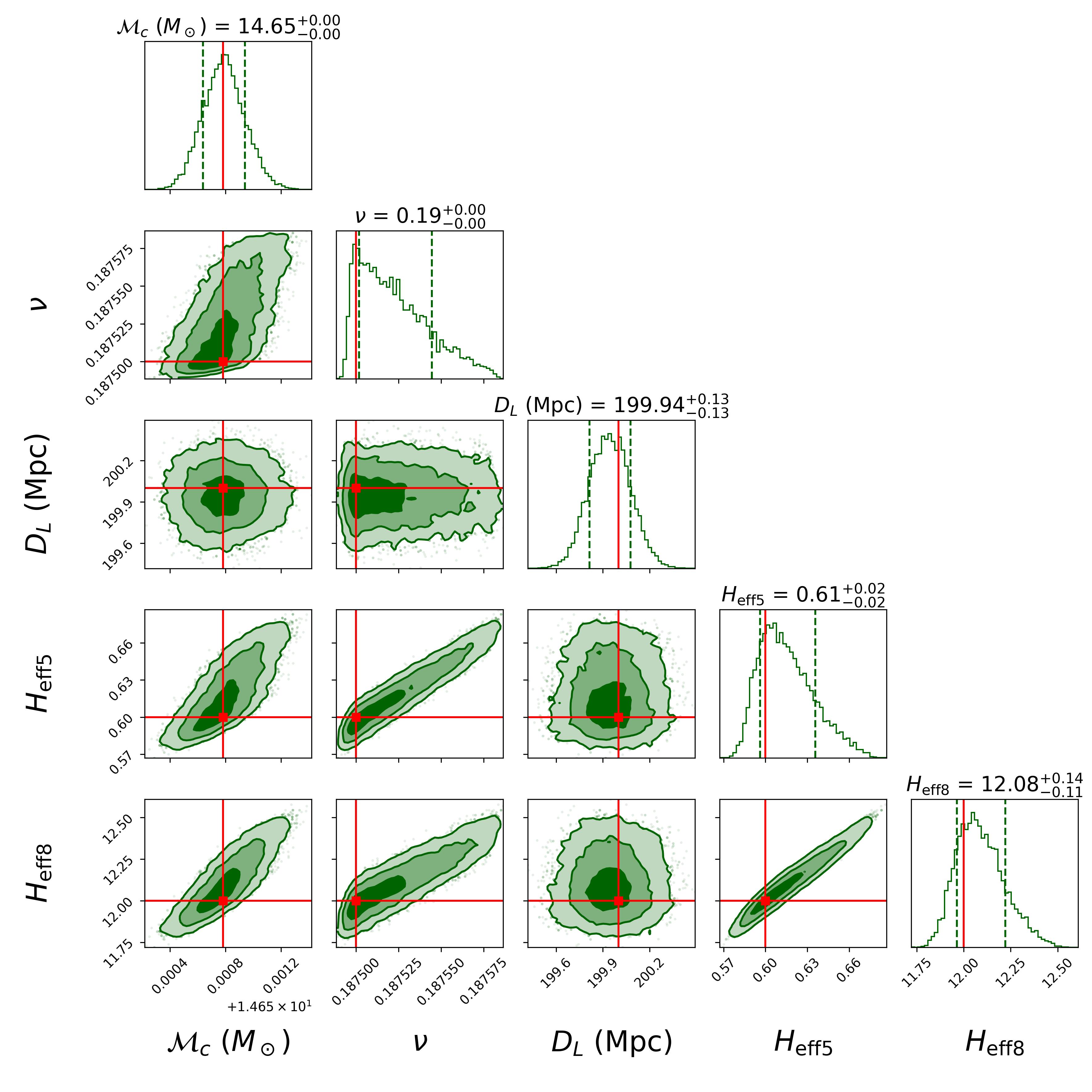}}
\subfigure[~Cosmic Explorer]{\label{40-ce}\includegraphics[width=0.48\textwidth]{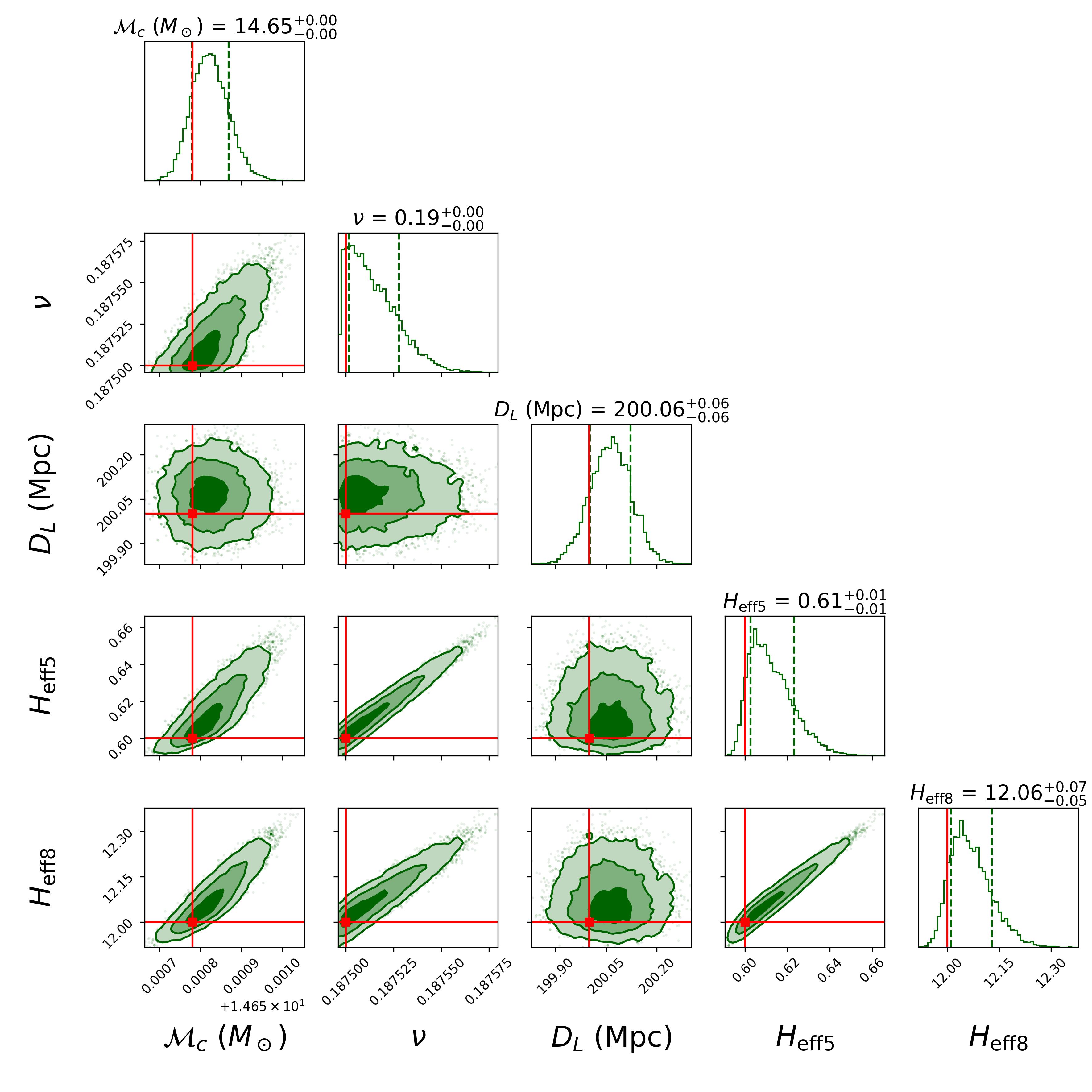}}

\caption{\small{Posterior plots from Bayesian parameter estimation. Injection parameters are $M=40 M_\odot$, $q=3$, $D^{}_L=200$ Mpc, $\chi^{}_1=\chi^{}_2=0.8$, $H_{\rm eff5}=0.6$, $H_{\rm eff8}=12$. The solid red lines denote the injected values of the corresponding parameters and the green dashed lines show the standard deviations (for the figures~\ref{40-et} and \ref{40-ce}, 14.65 is to be added to the tick labels of $\mathcal{M}_c$, which is mentioned below the x-axes of the subplots). The priors chosen for these simulations are listed in Table~\ref{priors}.}}
\label{bayes}
\end{figure*}

\subsubsection{Dependence on the Spins}
\label{sec:spin}

Measurements of $H_{\rm eff5}, H_{\rm eff8}$ are expected to depend on the spins of the binary components $\chi^{}_1,\chi^{}_2$ due to the presence of the spin-orbit term $\Psi^{}_{\rm SO}$ (Eq.~\eqref{spin-orbit}), and the fact that the upper cutoff frequency $f_{\rm ISCO}$ depends on the component spins (Eq.~\eqref{isco}). Figure~\ref{spinvar} shows contours of the error values in $H_{\rm eff5}$ and $H_{\rm eff8}$ in ET and CE detectors when the dimensionless (aligned) spins $\chi_1, \chi_2$ are varied from 0 to 1. $\Delta H_{\rm eff5}$ and $\Delta H_{\rm eff8}$ contours are shown in the plots in the left and the right columns, respectively. The parameter space considered for this analysis is $\Theta \equiv\{\mathcal{M}_c,\eta,D_L,\chi_1,\chi_2,H_{\rm eff5},H_{\rm eff8}\}$.
We demonstrate the spin dependence of the errors for total binary mass $M=40M_\odot$ and two values of the mass-ratios, $q=1.1$ (top panel) and $q=3$ (bottom panel). Let us consider one of these binaries in CE, with $q=1.1$ and low values of component spins, $\chi_1=\chi_2=0.2$. For the 7-dimensional parameter space mentioned above, the errors in $H_{\rm eff5}$ ($H_{\rm eff8}$) are $\sim 0.4$ (3), which amounts to a percentage error of $\sim 67\%$ ($25\%$) for this binary.   
The contours have lower error values as they approach the point $\chi_1=\chi_2=1$, indicating that the errors decrease with increasing spins. This can be attributed to the fact that $f_{\rm ISCO}$ increases with $\chi_1$ and/or $\chi_2$, making the effective frequency range larger, consequently adding more GW cycles in the frequency band. We note here that the values of the parameters themselves increase with the (aligned) spins substantially (Eq.~\eqref{Eq.Hparams}; Fig. 1 and Fig. 2 of Ref.~\cite{datta2020recognizing}), implying that for highly spinning compact objects one can put more stringent constraints on them.

\section{comparison with bayesian analyses}
\label{sec:bayesian}

We carried out Bayesian parameter estimation with  \texttt{Bilby}~\cite{Ashton:2018jfp} to compare the results with ones gotten from the Fisher analyses. For each of the detector networks (LIGO-Virgo, ET, CE), we chose one point from the region of the parameter space that is expected to produce the best results according to the Fisher studies above. This was partly to ensure the robustness of the best estimates found by the latter method. Although these regions are different for LIGO-Virgo and 3G detectors, as noted earlier, we choose the values of total mass ($M=40M_\odot$) and mass-ratio ($q=3$) same for all three detector networks for the sake of comparison. We first inject \texttt{TaylorF2} waveforms with the TH phase, then run the parameter estimation to obtain posteriors from the simulations. The starting frequency is 10Hz(4Hz) for LIGO-Virgo(ET, CE), and the upper cutoff frequency is taken to be  the corresponding ISCO frequency. 

In Table~\ref{priors}, we list the distribution and ranges of priors used for the chosen parameter space.


\begin{table}[h]
\centering
\begin{tabular}{|c c c c|} 
 \hline
 Parameter & Distribution & Range & Units \\ [0.5ex] 
 \hline\hline
 \small{Chirp mass} ($\mathcal{M}_c$) & Uniform & (10, 20) & $M_\odot$ \\ 
 \hline
 \small{Symmetric mass-ratio} & & & 
 \\ ($\nu$) & Uniform & (0.01, 0.25) & $\cdots$ \\
 \hline
 \small{Luminosity distance} & & & \\ 
 ($D_L$) & Uniform & (100, 500) & Mpc \\
 \hline
 $H_{\rm eff5}$ & Uniform & (-4, 4) & $\cdots$ \\
 \hline
 $H_{\rm eff8}$ & Uniform & (-20, 20) & $\cdots$ \\ [1ex] 
 \hline
\end{tabular}
\caption{\small Choice of priors for the Bayesian posteriors presented in Fig.~\ref{bayes}.}
\label{priors}
\end{table}

Figure~\ref{bayes} shows the corner plots generated form the posteriors, for the three detector-networks, LIGO-Virgo (Fig.~\ref{40-ligo}), ET (Fig.~\ref{40-et}), and CE (Fig.~\ref{40-ce}). As expected, estimation of $H_{\rm eff5}$ and $H_{\rm eff8}$ are much better in ET and CE than in LIGO-Virgo, with the errors broadly agreeing with their Fisher counterparts for similar systems studied in Figs.~\ref{h5-ligo-m},~\ref{h5-et-m},~\ref{h5-ce-m} for $H_{\rm eff5}$, and Figs.~\ref{h8-ligo-m},~\ref{h8-et-m},~\ref{h8-ce-m} for $H_{\rm eff8}$.

\begin{figure}[]
    \centering
    \includegraphics[width=86mm]{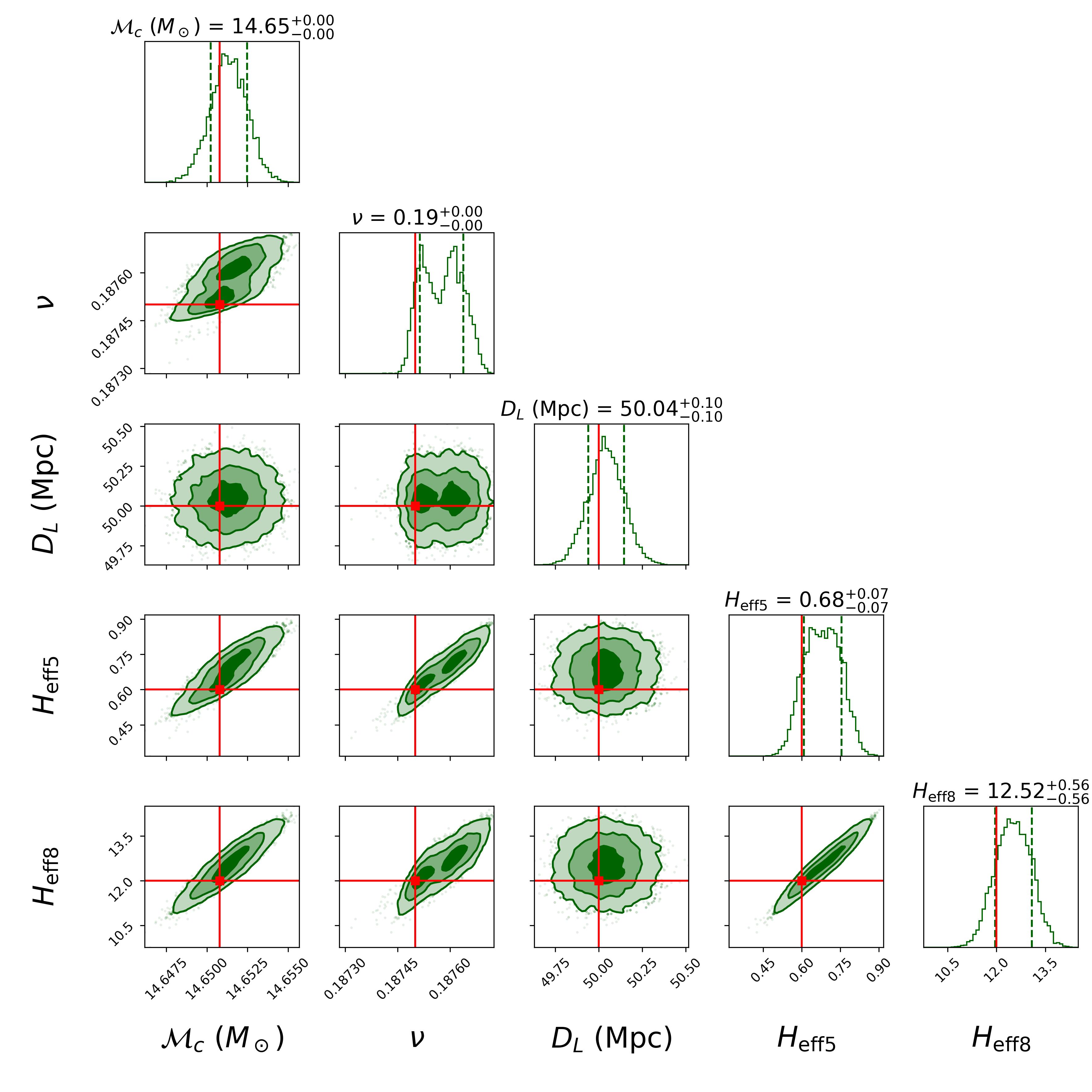}
    \caption{\small Bayesian posterior plot in LIGO-Virgo \\ with $M=40 M_\odot, q=3, D_L=50$ Mpc, $\chi^{}_1=\chi^{}_2=0.8$, $H_{\rm eff5}=0.6$, $H_{\rm eff8}=12$.}
    \label{ligo-golden}
\end{figure}

We chose a fourth point for the LIGO-Virgo network, which is significantly close -- at $D_L=50$ Mpc, with all the other parameters the same as in Fig.~\ref{40-ligo}. Prior for the luminosity distance is taken to be uniform in the range (10, 100) Mpc. All the other parameters have the same priors as in Table~\ref{priors}. Figure~\ref{ligo-golden} shows the corresponding posterior plot. Comparing 
the two, 
we see the expected improvement in accuracy and precision, arising from the binary being four times closer. The errors in this case are $\sim 11.7\%(\sim 4.7\%)$ for $H_{\rm eff5} (H_{\rm eff8})$ for a BBH.


\begin{figure*}[ht]
    \centering
    \hspace{-1cm}
    \includegraphics[width=\textwidth]{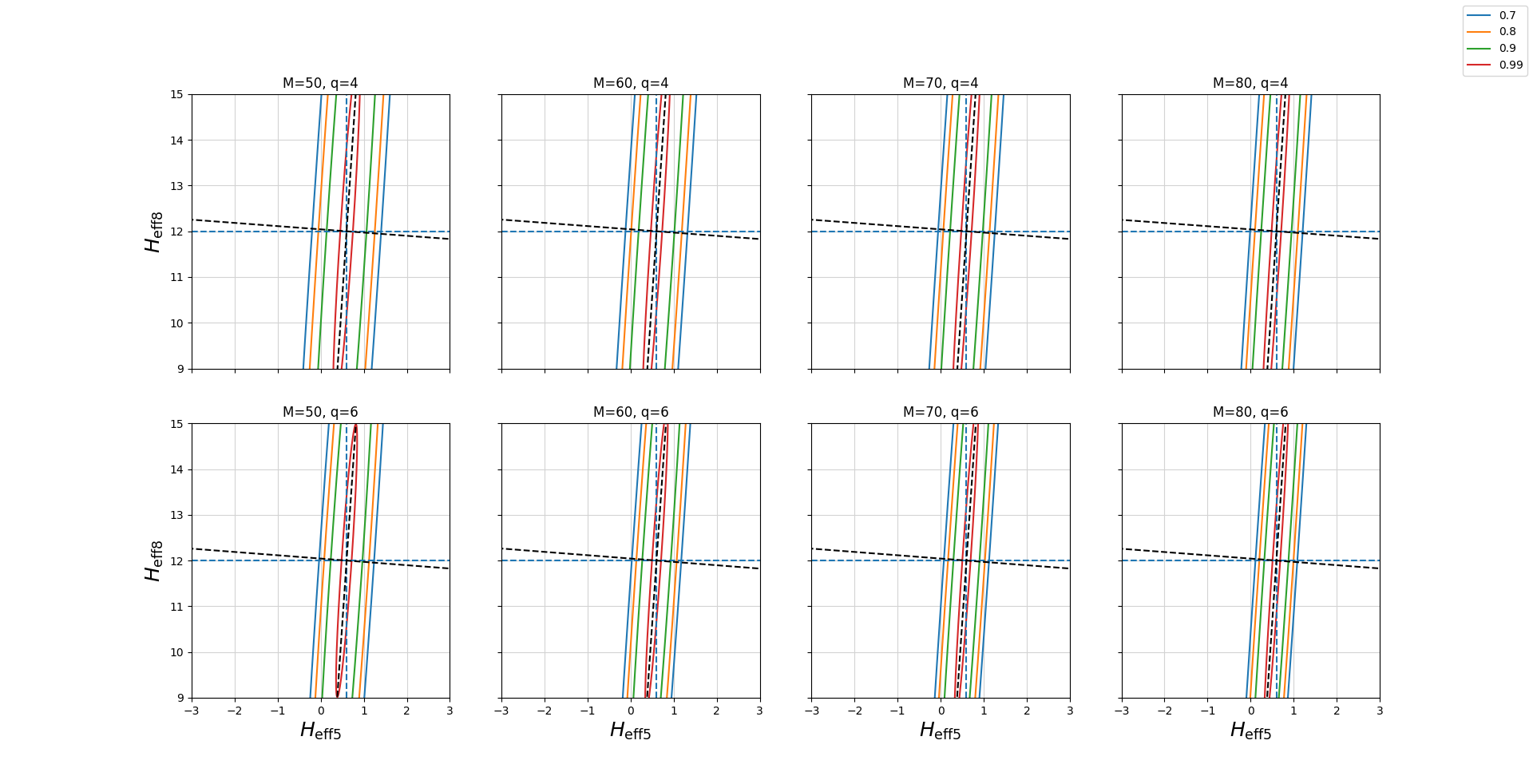}
    \caption{\small Fisher ellipses in LIGO-Virgo with normalized templates, implying SNR = 1. The spin values taken here are $\chi_1=\chi_2=0.8$. The ``target waveform" corresponds to the intersection point of the eigenvector directions (the dotted black lines), which is $H_{\rm eff5}=0.6, H_{\rm eff8}=12$. First and second row correspond to mass-ratio $q=4$ and $q=6$ respectively, and the four columns are for total mass values $M=50,60,70,80M_\odot$ from left to right. Match values for different ellipses are shown in the common legend at the upper right corner.}
    \label{ellipse_ligo}
\end{figure*}


\begin{figure*}[]
    \hspace{-1cm}
    \subfigure[]{\label{ellipse-et}\includegraphics[width=\textwidth]{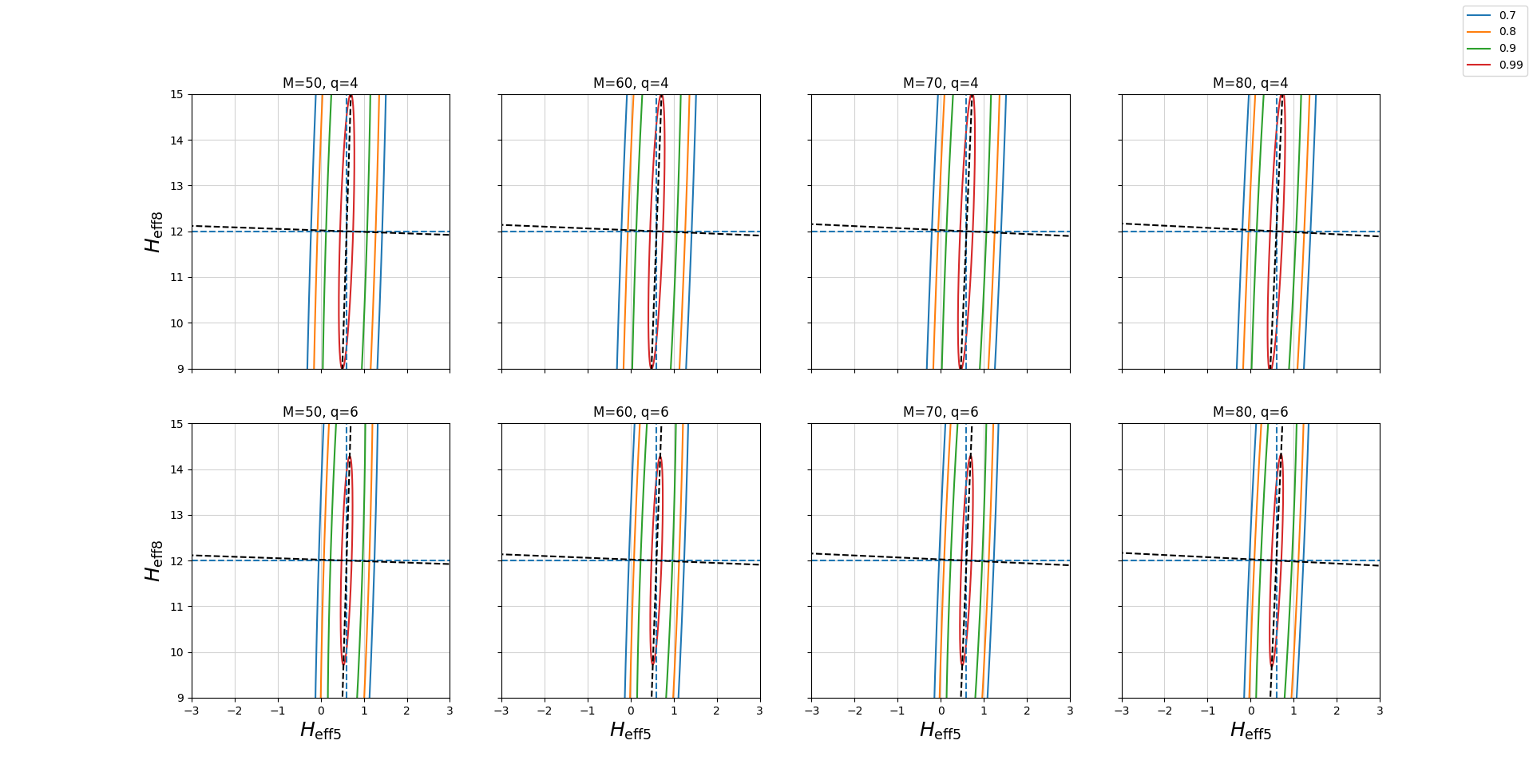}}
    \newline
    \hspace*{-1.3cm}
    \subfigure[]{\label{ellipse-ce}\includegraphics[width=\textwidth]{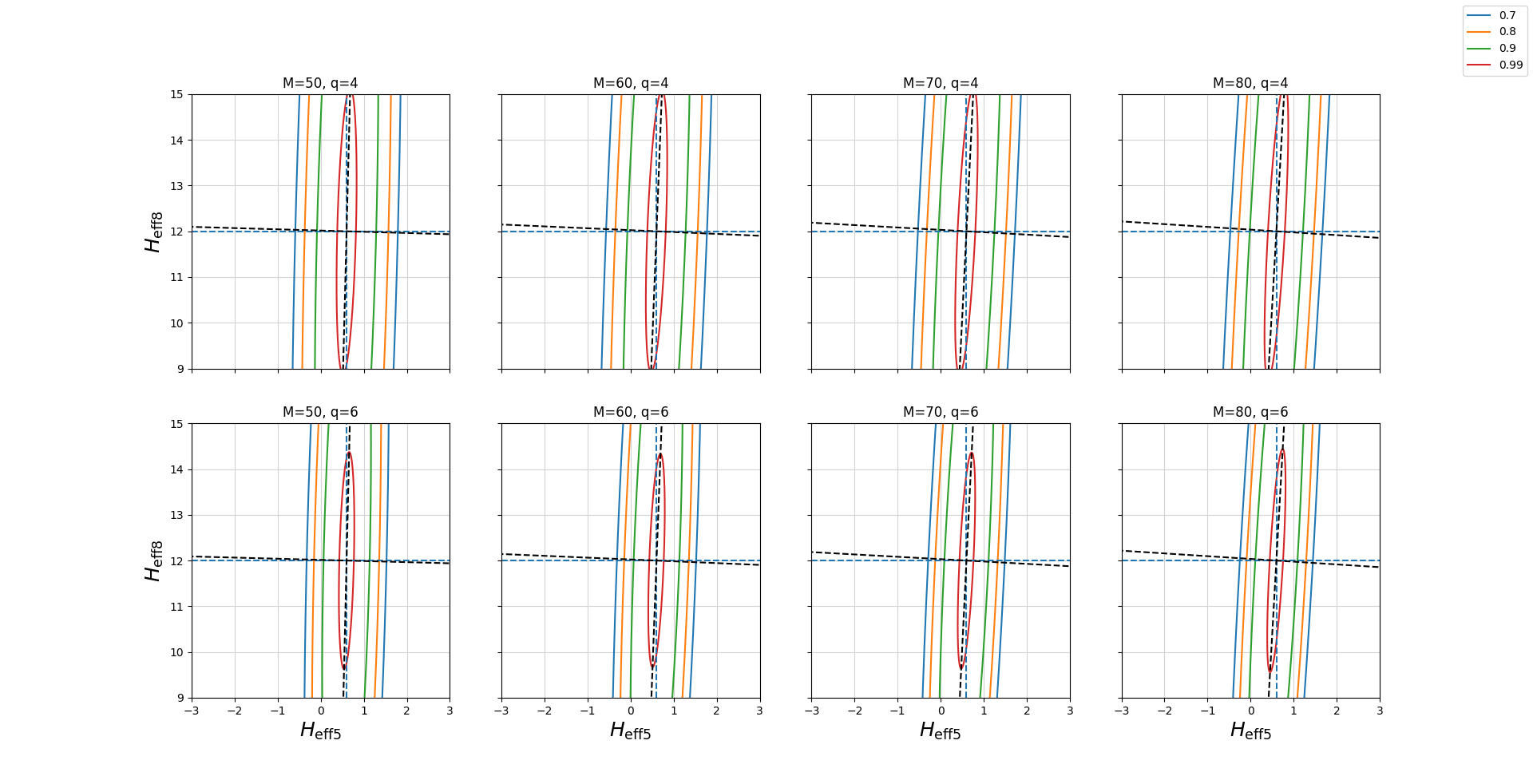}}
    
    \caption{\small Same as in Fig.~\ref{ellipse_ligo}, but in the detectors ET (Fig.~\ref{ellipse-et}) and CE (Fig.~\ref{ellipse-ce}). }
    \label{ellipse_et_ce}
\end{figure*}


\section{principal component analysis}
\label{pca}

So far we have considered the diagonal elements of the error
covariance matrix $C$, which are the variances. In this section we will consider the covariances between $H_{\rm eff5}$ and $H_{\rm eff8}$, which are the off-diagonal elements of $C$. Diagonalizing $C$ will  yield its eigenvalues and eigenvectors. The eigenvectors will provide the new coordinates (i.e., horizon parameters) that have a vanishing  covariance, and the eigenvalues will correspond to the errors in those new coordinates. Keeping this in mind, we choose a particular waveform as a ``target" waveform, with which we calculate the matches (defined below) of several neighboring ``template" waveforms. If we make a contour plot of these match values, we get ellipses -- with their principal axes along the eigenvectors of $C$. Since $C=\Gamma^{-1}$, the eigenvectors of $C$ and $\Gamma$ are the same, and their eigenvalues are related by $\lambda_i=1/\alpha_i$, where $\lambda_i(\alpha_i)$ are the eigenvalues of $\Gamma(C)$.

In this quest, we take the manifold $\mathcal{V}$ of dimensionality same as that of $\Theta$, with every point on $\mathcal{V}$ corresponding to a template waveform. The distance between two templates on $\mathcal{V}$ corresponding to the parameter vectors $\Theta$ and $(\Theta+\Delta\Theta)$ can be calculated as~\cite{Toubiana:2020cqv}:

\begin{align}
\label{temp_distance}
    |h(\Theta+\Delta\Theta)&-h(\Theta)|^2 \nonumber\\
    =&\braket{h(\Theta+\Delta\Theta)-h(\Theta)}\nonumber\\
    =&\braket{\pdv{h}{\Theta_i}}{\pdv{h}{\Theta_j}}\Delta\Theta^i\Delta\Theta^j\nonumber\\
    =&\Gamma_{ij}\Delta\Theta^i\Delta\Theta^j.
\end{align}
The \textit{match} ($\mathcal{M}$), also known as the \textit{overlap function}~\cite{Ajith:2009fz,owen_PhysRevD.53.6749}, between two template waveforms $h(\Theta)$ and $h(\Theta+\Delta\Theta)$ can be defined as the inner product between them, maximized over the extrinsic parameters $t_c$ and $\phi_c$~\cite{owen_PhysRevD.53.6749} :
\begin{equation}
    \mathcal{M} = \underset{t_c,\phi_c}{\rm max}\braket{h(\Theta+\Delta\Theta)}{h(\Theta)}.
\end{equation}

If all the templates $h(\Theta)$ on $\mathcal{V}$ are normalized to $\hat{h}(\Theta)$ by $\hat{h}(\Theta) = h(\Theta)/\braket{h(\Theta)}{h(\Theta)}$, so that $\braket{\hat{h}(\Theta)}{\hat{h}(\Theta)}=1\, \forall\, \hat{h}(\Theta)\in \mathcal{V}$, then the maximum value of $\mathcal{M}$ can be 1, which corresponds to $\Delta\Theta=0$. Then, one can define the \textit{mismatch} between two templates $\hat{h}(\Theta)$ and $\hat{h}(\Theta+\Delta\Theta)$ as $1-\mathcal{M}$, which geometrically denotes the ``distance" between them on the manifold $\mathcal{V}$. One can relate them, using Eq.~\eqref{temp_distance}, as
\begin{equation}
\label{match}
    1-\mathcal{M}=\Gamma^{(n)}_{ij}\Delta\Theta^i\Delta\Theta^j.
\end{equation}

Here $\Gamma^{(n)}_{ij}$ is the Fisher matrix for normalized templates ($\hat{h}(\Theta)$), related to the Fisher matrix for unnormalized templates ($h(\Theta)$) as 

\begin{align}
    \Gamma_{ij}=&\braket{\pdv{h}{\Theta_i}}{\pdv{h}{\Theta_j}}\nonumber\\
    =&\braket{h}{h}\braket{\pdv{\hat{h}}{\Theta_i}}{\pdv{\hat{h}}{\Theta_j}}\nonumber\\
    =&\,\rho^2\, \Gamma^{(n)}_{ij}.
\end{align}
The last expression follows from the fact that $\sqrt{\braket{h}{h}}$ is the SNR $\rho$, given by Eq.~\eqref{snr2}.
Eq.~\eqref{match} motivates one to define a metric $g_{ij}$ on $\mathcal{V}$ to express the distance between two templates $\hat{h}(\Theta)$ and $\hat{h}(\Theta+\Delta\Theta)$ as $g_{ij}\Delta\Theta^i\Delta\Theta^j$, and identify the relation of the metric with the Fisher matrix as $g_{ij}=\Gamma^{(n)}_{ij}=(1/\rho^2)\Gamma_{ij}$.

In our analysis, we consider a 2D manifold with only $H_{\rm eff5}$ and $H_{\rm eff8}$ as parameters, which is a submanifold of $\mathcal{V}$ with all the other parameters fixed. On this submanifold, Eq.~\eqref{match} can be expanded as
\begin{equation}
\begin{split}
    1-\mathcal{M}=\Gamma^{(n)}_{00}(H_{\rm eff5}-H_{\rm eff5}^\ast)^2+\Gamma^{(n)}_{11}(H_{\rm eff8}-H_{\rm eff8}^\ast)^2\\+2\Gamma^{(n)}_{01}(H_{\rm eff5}-H_{\rm eff5}^\ast)(H_{\rm eff8}-H_{\rm eff8}^\ast)\,,
\end{split}
\end{equation}
with $H_{\rm eff5}^\ast(H_{\rm eff8}^\ast)$ being the value of $H_{\rm eff5}(H_{\rm eff8})$ corresponding to the target waveform.
Thereby, the contours of constant values of $\mathcal{M}$ represent ellipses in the space of $H_{\rm eff5}$ and $H_{\rm eff8}$, centered at ($H_{\rm eff5}^\ast$,$H_{\rm eff8}^\ast$), given that the Fisher matrix components are constant.
For a Fisher matrix with only $H_{\rm eff5}$ and $H_{\rm eff8}$ as parameters, none of its components depends on the values of $H_{\rm eff5}$ or $H_{\rm eff8}$. This implies that the metric is flat on this submanifold, and the contours of constant $\mathcal{M}$ are all perfect ellipses. If we diagonalize $\Gamma$, then in the eigen-coordinates, all the covariances will vanish. Let us call the corresponding eigenvectors $(X,Y)$, known as the {\it principal components}~\cite{Ohme:2013nsa}. Then the equation of the ellipses with respect to the eigen-coordinates $(X,Y)$ with corresponding eigenvalues ($\lambda_1, \lambda_2$) becomes
    \begin{equation}
        \lambda_1 (X-X^\ast)^2+\lambda_2 (Y-Y^\ast)^2=(1-\mathcal{M}).
    \end{equation}
These ellipses are centered at $(X^\ast,Y^\ast)$, and have principal axes ($a,b$) given by,
\begin{equation}
    a=\sqrt{(1-\mathcal{M})/\lambda_1}\,, \quad\quad b=\sqrt{(1-\mathcal{M})/\lambda_2}\,.
\label{a,b}
\end{equation}
 
Figure~\ref{ellipse_ligo} shows such ellipses in the sensitivity of Advanced LIGO and Virgo for a target waveform with source parameters $H_{\rm eff5}^\ast=0.6, H_{\rm eff8}^\ast=12$. Figures~\ref{ellipse-et} and \ref{ellipse-ce} show similar ellipses in ET and CE, respectively. We show eight different plots for a combination of different values of the total mass and mass-ratio. Since $\Gamma$ is a symmetric matrix, its eigenvectors, lying along the dotted lines shown in the plots, are orthogonal to each other. 

\begin{figure}[]
    \centering
    \includegraphics[width=85mm]{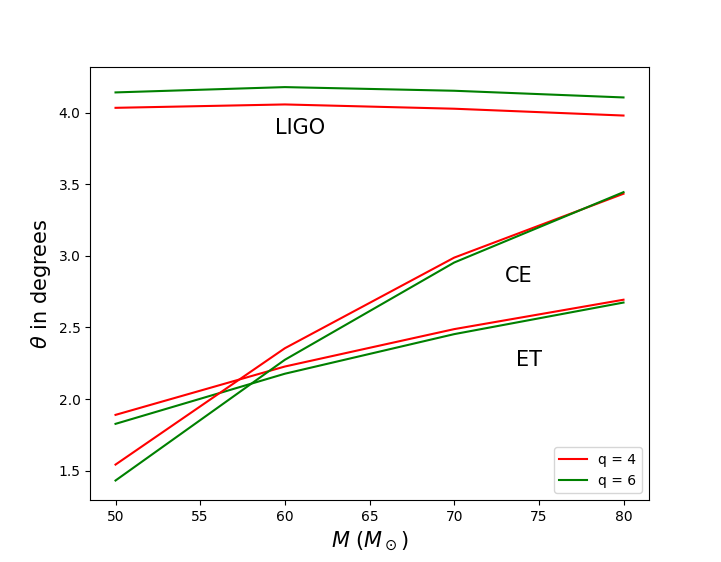}
    \caption{\small Variation of the rotation angle of $X-Y$ coordinate axes with respect to the $H_{\rm eff5}-H_{\rm eff8}$ axes with $M$ for $q=4,6$ in Fig.~\ref{ellipse_ligo} and Fig.~\ref{ellipse_et_ce}. }
\label{theta}
\end{figure}

 Covariances between the two parameters cause the ellipses to tilt, with higher tilt angles implying higher covariances between $H_{\rm eff5}$ and $H_{\rm eff8}$.  In Fig.~\ref{theta}, we show the variation of the tilt angles ($\theta$) between the $X-Y$ coordinate axes and the $H_{\rm eff5}-H_{\rm eff8}$ axes with total mass $M$, for $q=4$ and $q=6$. In LIGO-Virgo, the tilt angles vary negligibly with $M$, but their values are higher compared to the 3G detectors. In ET and CE, The tilts of the ellipses increase slowly with $M$, implying that the covarinaces between $H_{\rm eff5}$ and $H_{\rm eff8}$ are higher for higher-mass systems. We also note that CE shows a faster growth in the covariances for higher mass systems than ET. The effect of $q$ on the covariances appears to be different in LIGO-Virgo than in ET, CE -- in the former, they increase with increasing $q$, but the latter two follow the opposite trend. The small tilt angles of the eigen-coordinates, especially in 3G detectors, imply negligible covariances between $H_{\rm eff5}$ and $H_{\rm eff8}$.
 
\begin{figure}[]
    \centering
    \includegraphics[width=85mm]{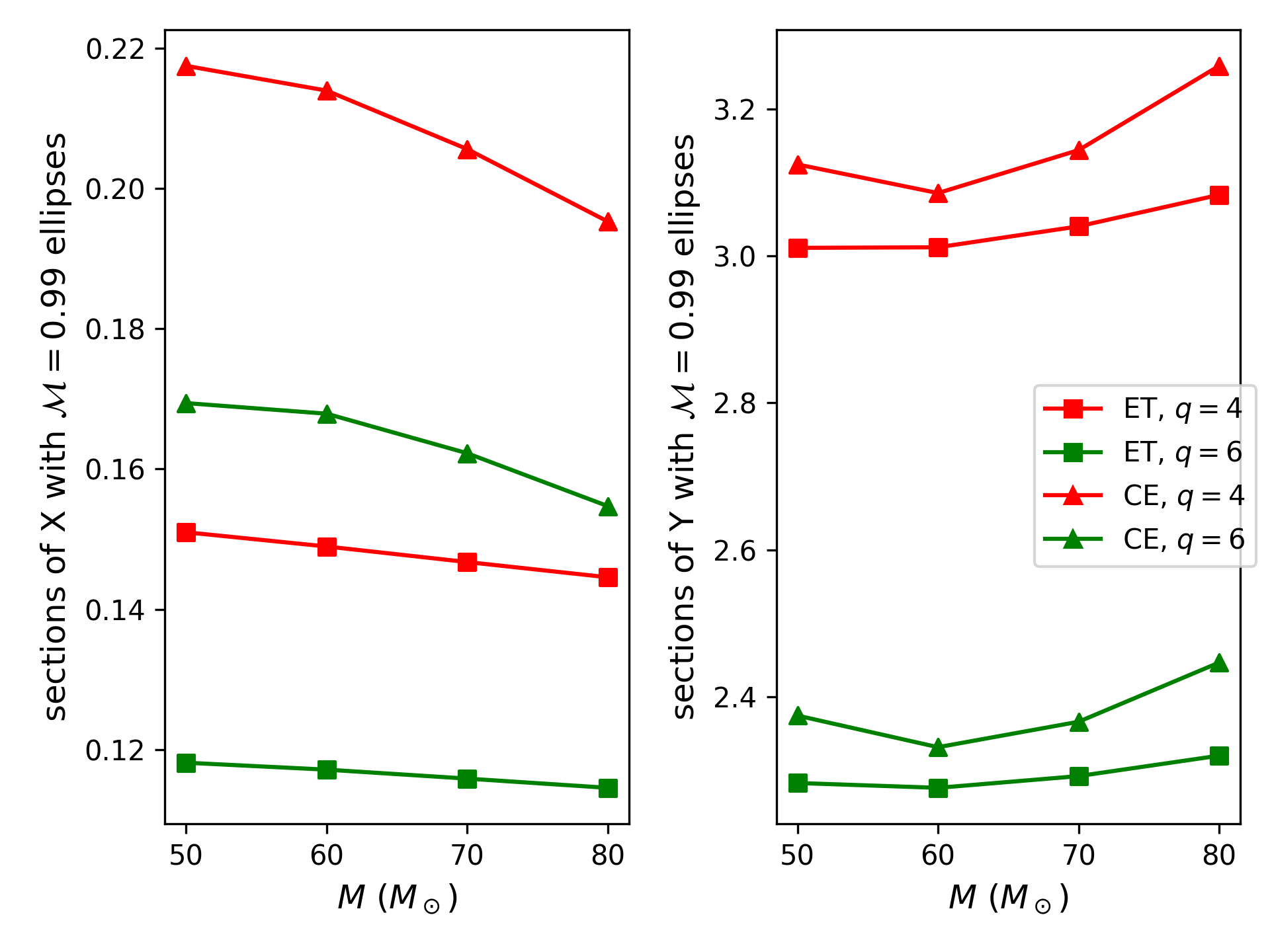}
    \caption{\small Sections of the $\mathcal{M}=0.99$ ellipses with their principal axes in Fig.~\ref{ellipse_et_ce}, plotted against $M$ for $q=4$ (red), $q=6$ (green) in ET (square points) and CE (triangular points).}
\label{eigval}
\end{figure}

 The measurability of a certain parameter can be inferred from these ellipses by studying how closely spaced they are along the direction of that parameter, which denotes how rapidly the match values change with small displacements along that direction. Rapid change of match values implies that two different waveforms can be distinguished better; consequently, the errors are smaller. Since we are considering only normalized waveforms for this analysis, effects of the SNR on the statistical errors are absent here, in contrast to Sec.~\ref{results} where the results depend largely on SNR. This enables us to study the variations of the errors in the eigen-coordinates in an SNR-independent way. To demonstrate how the shapes of the ellipses vary with total mass and mass-ratio, in Fig.~\ref{eigval} we plot the principal axes $a$ and $b$ of the ellipses, defined in Eq.~\eqref{a,b}, for the match value $\mathcal{M}=0.99$ (the red ellipses in Fig.~\ref{ellipse_ligo} and Fig.~\ref{ellipse_et_ce}). In this figure only the 3G detectors are considered. The ellipses get stretched out along the eigen-coordinate $Y$ (the semi-major axes) with increasing $M$, implying that the error in that coordinate increases with $M$ for $M> 60M_\odot$. This follows the behavior of $\Delta H_{\rm eff8}$, which increases with $M$ in this region (Figs.~\ref{h8-et-m},~\ref{h8-ce-m}). Along the $X$ direction (semi-minor axes), sections of the ellipses are smaller with increasing $M$,
 implying lesser errors.  Increasing $q$ makes the ellipses more squeezed along both $X$ and $Y$, implying better measurabilities  in both the eigen-coordinates.


Covariance between two parameters is a measure of the degeneracy between them. A vanishing covariance between two parameters allows one to probe the dependence of the model (in our case, the gravitational waveform) on them separately. This possibility is attained by 3G detectors with the introduction of more GW cycles in the early inspiral, and the accompanying fact that these parameters are much more precisely measurable than in LIGO and Virgo. In the case of ET and CE, as we see, in the considered region of the parameter space, the errors and the covariances are not large enough for diagonalization to make any significant difference.

\section{Discussion and Conclusions}
\label{discussion}

We have explored how well one can measure the two tidal heating parameters $H_{\rm eff5}$ and $H_{\rm eff8}$ in the future ground-based GW detectors Einstein Telescope and Cosmic Explorer as well as in the  2nd generation detectors Advanced LIGO and Advanced Virgo. These parameters account for the flux of energy and angular momentum into 
or out of a (spinning) BH, which is different for other compact objects -- even those that mimic a BH. They appear at 2.5PN and 4PN orders, respectively, in the expression for the phase shift in the gravitational waveform due to tidal heating. 
The prospect of proper estimation of these parameters results in a viable method for distinguishing BHs (in binaries) from other compact objects that do not have horizons, but may otherwise  resemble them. We  chose \texttt{TaylorF2} as the PN approximant and added the tidal-heating phase shift to it. We used primarily the Fisher matrix approach  for  estimating the errors. 

In 3G detectors, we showed that for a total binary mass of $M\gtrsim 50M_\odot$, estimation of the aforementioned parameters is the most precise, whereas for  2G detectors there is a specific region around $20M_\odot\lesssim M \lesssim 40M_\odot$ where we expect the best results with the current waveform. Increasing mass asymmetry results in lesser errors. The errors rise linearly with the luminosity distance. In 3G detectors, we can constrain $H_{\rm eff5}$ ($H_{\rm eff8}$) with a 1-$\sigma$ error value of $\sim 0.05$ ($\sim 0.2$) for a binary with $M=30M_\odot, q=1.5, H_{\rm eff5}=0.6, H_{\rm eff8}=12$, at 1 Gpc distance. This error value amounts to a relative percentage error of approximately 8.3\% (2\%) for $H_{\rm eff5}$ ($H_{\rm eff8}$). In LIGO-Virgo, the errors are higher ($\sim 300\%$ for $H_{\rm eff5}$, $\sim 50\%$ for $H_{\rm eff8}$), as expected, mainly due to the lower SNRs. The measurements can be improved by using coherent mode stacking, by which one can combine observations of $N$ number of GW events and effectively scale the SNR by a factor of $\sqrt{N}$~\cite{Yang:2017zxs,Bose:2017jvk}. Spins of the binary components affect the measurabilities due to the spin-orbit term, and the fact that the upper cutoff frequency used in this work is spin-dependent. Increasing spin makes the considered frequency range wider, which in turn lowers the error values.

We have also demonstrated that in the sensitivity of 3G detectors, covariances between these parameters are not significant, meaning that we do not expect to improve the results any further by introducing any new combination of them by diagonalizing the covariance matrix. However, in 2G detectors, covariances are high enough for this method to produce better results, and we show how we can define a new set of coordinates with less errors from the tilts of the Fisher ellipses. 

This work will be useful in future studies when more complete and accurate tidal-heating waveforms are available that extend deeper into the merger phase. Our study has identified the regions in the parameter space where one can expect the best results in estimating the tidal heating parameters. We have shown that these results are consistent with Bayesian analyses.

\section*{acknowledgments}

SM and SD would like to thank University Grants Commission (UGC), India, for financial support for a senior research fellowship. We thank Muhammed Saleem, Michalis Agathos, K.G. Arun and N.V. Krishnendu for helpful discussions and comments. This article has been assigned the LIGO document number LIGO-P2100474.

\bibliography{references}
\end{document}